\documentclass{aa}
\usepackage{lscape}
\usepackage{graphicx}
\usepackage{txfonts}
\usepackage{natbib}
\usepackage{lscape}
\bibpunct{(}{)}{;}{a}{}{,} 
\usepackage{deluxetable}

\newcommand{\pedix}[2]{\ensuremath{#1_{\mbox{\scriptsize #2}}}}
\newcommand{\apix}[2]{\ensuremath{#1^{\mbox{\scriptsize #2}}}}
\newcommand{\pedap}[3]{\ensuremath{#1_{\mbox{\scriptsize #2}}^{\mbox{\scriptsize #3}}}}
\newcommand{\pedSM}[2]{\ensuremath{#1_{\mbox{\tiny #2}}}}

\newcommand{\errUD}[2]{\ensuremath{^{+#1}_{-#2}}}

\newcommand{\mydag}{\ensuremath{\mbox{\dag}}}
\newcommand{\myddag}{\ensuremath{\mbox{\ddag}}}

\newcommand{\tildetxt}[1]{\ensuremath{\tilde{\mbox{#1}}}}
\newcommand{\ntilde}{\tildetxt{n}}

\newcommand{\xmm}{{XMM-\emph{Newton}}}

\newcommand{\iue}{\emph{IUE}}
\newcommand{\hst}{\emph{HST}}
\newcommand{\rosat}{{\emph{ROSAT}}}
\newcommand{\asca}{{\emph{ASCA}}}
\newcommand{\chandra}{{\emph{Chandra}}}

\newcommand{\sdss}{{\emph{SDSS}}}

\newcommand{\lum}{\ensuremath{\mbox{ergs~s}^{-1}}}
\newcommand{\flux}{\ensuremath{\mbox{ergs~cm}^{-2}\mbox{~s}^{-1}}}
\newcommand{\fluxA}{\ensuremath{\mbox{ergs~cm}^{-2}\mbox{~s}^{-1}\mbox{~\AA}^{-1}}}
\newcommand{\nh}{\ensuremath{\mbox{cm}^{-2}}}
\newcommand{\nhsym}{\ensuremath{N_{\mbox{\scriptsize H}}}}
\newcommand{\arcdeg}{\ensuremath{^{\circ}}}
\renewcommand{\arcmin}{\ensuremath{^{\prime}}}
\renewcommand{\arcsec}{\ensuremath{^{\prime\prime}}}
\newcommand{\kev}{\ensuremath{\,\mbox{\scriptsize keV}}}
\newcommand{\ang}{\ensuremath{\,\mbox{\scriptsize \AA}}}

\newcommand{\normPL}{\ensuremath{\mbox{photons~keV}^{-1}\mbox{~cm}^{-2}\mbox{~s}^{-1}}}

\newcommand{\chidof}{\ensuremath{\chi^2/\mbox{d.o.f.}}}
\newcommand{\OIII}{[\ion{O}{iii}]}   
\newcommand{\hb}{\ensuremath{\mbox{H}\beta}}
\newcommand{\ha}{\ensuremath{\mbox{H}\alpha}}
\newcommand{\feka}{\ensuremath{\mbox{Fe~K}\alpha}}

\newcommand{\tbox}[1]{\mbox{\tiny #1}}
\def\text{\tbox}
\newcommand{\ie}{{\emph{i.e.}}}
\newcommand{\eg}{{\emph{e.g.}}}

\newcommand{\sorg}{\object{PG~1535+547}}
\newcommand{\nhgal}{\ensuremath{1.35\times10^{20}}}
\newcommand{\redgal}{\ensuremath{0.038}}
\newcommand{\posgal}{RA$=$15$:$36$:$38.3, Dec$=+$54$:$33$:$33}
\newcommand{\dataO}{November~2002}
\newcommand{\dataN}{January~2006}

\begin{document}

\title{X-ray spectral variability in PG~1535+547: the changing-look of a ``soft X-ray weak'' AGN}

\author{
	L.~Ballo\inst{1},
	M.~Giustini\inst{2,}\inst{3},
	N.~Schartel\inst{1},
	M.~Cappi\inst{3},
	E.~Jim\'enez-Bail\'on\inst{4},
	E.~Piconcelli\inst{5},
	M. Santos-Lle{\'o}\inst{1},
	\and
	C.~Vignali\inst{2,}\inst{6}
	}

\offprints{L.~Ballo, \email{lucia.ballo@sciops.esa.int}}

\institute{\inst{1}\xmm\ Science Operation Centre, ESAC, ESA, P.O. Box 78, E-28691 Villanueva de la Ca{\ntilde}ada, Madrid, Spain \\
           \inst{2}Dipartimento di Astronomia, Universit\`a degli Studi di Bologna, via Ranzani 1, I-40127, Bologna, Italy \\
           \inst{3}Istituto di Astrofisica Spaziale e Fisica cosmica (INAF), via Gobetti 101, I-40129, Bologna, Italy \\
           \inst{4}Instituto de Astronom\'\i{}a, Universidad Nacional Aut\'onoma de M\'exico, Apartado Postal 70-264, 04510-Mexico DF, M\'exico \\
           \inst{5}Osservatorio Astronomico di Roma (INAF), via Frascati 33, I-00040 Monteporzio Catone, Roma, Italy \\
           \inst{6}Osservatorio Astronomico di Bologna (INAF), via Ranzani 1, I-40127 Bologna, Italy}

\date{Received XXXXXXXXXX XX, XXXX; accepted XXXXXXXXXX XX, XXXX}

\abstract
 {\sorg\ is a bright Narrow Line Seyfert~1 galaxy, whose high energy emission shows strong variability both in shape and flux.
  On the basis of \rosat\ observations, it is classified as ``soft X-ray weak QSO'', a class of objects whose X-ray--to--optical flux ratio is smaller than in typical QSOs.
   Their X-ray spectra are often characterized by highly ionized, complex absorbers and/or reflection from the inner accretion disk, and the relative importance of the two is currently debated.
  Whatever the correct interpretation may be, the presence of such features implies that we are looking at matter located in the innermost regions of these AGN.}
 {We want to clarify the nature of the X-ray emission of \sorg, and constrain the physical properties of its innermost regions, where this emission originates.}
 {We present new \xmm\ observations of \sorg\ ($90\;$ksec exposure time), from which we obtained two spectra separated by about one week, that we compare with a previous (about three years) \xmm\ observation.}
 {These observations support the complex and variable nature of the X-ray emission of \sorg.
  The broad band observed flux increases by a factor $\sim 2.3$ in three years, and then decreases by a factor $\sim 1.3$ in about one week.
  In the new EPIC spectra strong absorption features at $E < 3\;$keV and a complex spectral shape in the iron line energy range are evident, coupled with a drop in the emission at higher energies.
  We describe all the different states in a consistent way, assuming either a warm absorber plus a relativistically blurred ionized reflection, or a two-phase warm absorber partially covering the source with the addition of a scattered component.}
 {The observed variability can be ascribed mostly to warm absorbing gas in the innermost regions of \sorg, that appears to vary in its physical properties on timescales of both years and days.
  In the blurred reflection scenario all the analysed states require a high fraction of reflection from the disk, calling for some mechanisms able to increase the reflection component with respect to the intrinsic continuum.
  Finally, the strong variability observed in the X-ray band opposed to a more constant emission at optical frequencies changes the value of the X-ray--to--optical spectral index, implying that \sorg\ can not actually be classified as a soft X-ray weak AGN.}


\titlerunning{X-ray variability in PG~1535+547}
\authorrunning{L.~Ballo et al.}

\maketitle


\section{Introduction}\label{sect:intro}

Studying the X-ray emission from Active Galactic Nuclei (AGN) offers the potential of investigate the central engine powering these sources.
The fluorescent Fe emission line complex at $6.4-7\;$keV is an important direct probe of the dense matter around the nuclear region, from the molecular torus and the Broad Line Region (BLR) to the inner accretion disk, down to a few gravitational radii ($\pedix{r}{g} \equiv G\pedix{M}{BH}/c^2$). 
In the latter case, intrinsically narrow emission lines emitted by the accretion disk are predicted to be broadened by Doppler and relativistic effects.
Detailed studies of their energy profiles and flux variability could give strong constraints on the geometry and kinematics of the accretion disk in terms of inclination, inner and outer radii of the emitting region and ionization state (\eg, \citealt{miller06}; see \citealt{reynolds03} for a review).

The inner region of AGN houses also ``warm'' (\ie, partially photoionized, $T\sim 10^5\;$K) gas that absorbs the nuclear emission in the X-ray band.
This ``warm absorber'' \citep{halpern84} is often found to be complex and multi-phase, and outflowing with velocities in the range $0.001-0.01\:c$ \citep[\eg,][]{piconcelli04,ashton04,blustin05,mckernan07}.
Recently, more extreme outflow velocities (\ie, up to $\sim 0.3\:c$) have been detected in the X-ray spectra of a few dozen of AGN \citep[for a review, see][]{cappi06}.
The knowledge of the properties of warm absorbers such as their location, covering fraction, mass content and outflow velocity can help us to understand not only the accretion/ejection mode of AGN, but also their impact on the surrounding medium.
Constraining the physical properties of ionized gas harbored in the inner regions of AGN and disentangle reflection from absorption contribution are challenges that can be unfolded only via high energy, temporally resolved spectral analyses.

Linked to both these topics, the origin of the so-called ``soft excess'', an enhancement of flux seen below $2-3\;$keV with respect to the extrapolation at low energies of the hard nuclear component, is still a debated point.
While the constancy  of its ``temperature'' over a wide range of luminosities and black hole masses seems in contrast with the idea of direct thermal emission from the accretion disk \citep[see \eg][]{piconcelli05}, two different explanations have been proposed \citep[both probably playing a fundamental role in a complex combination, see \eg][]{chevallier06}: photoionized reflection from the inner accretion disk \citep{crummy06}, or absorption from a relativistically outflowing, moderately ionized medium \citep{swind04,swind06,schurch06,sobolewska07}.

Over the past years, a comparison between optical and X-ray properties of bright sources pointed out the existence of a population of AGN notably faint in soft X-ray relative to their optical fluxes, the so-called ``soft X-ray weak QSO'' \citep{laor97}.
Analysing the X-ray emission of the \citet{boroson92} sample of nearby AGN, \citet{brandt00} found that approximately $10$\% of optically selected AGN are X-ray weak at soft energies, with an X-ray--to--optical spectral index $\pedix{\alpha}{ox} \equiv \log (\pedix{F}{2\kev}/\pedix{F}{3000\ang})/\log (\pedix{\nu}{2\kev}/\pedix{\nu}{3000\ang}) < -2$.
The strong correlation observed between \pedix{\alpha}{ox} and absorption features in the UV band (mainly the \ion{C}{iv} absorption equivalent width) supports the idea of X-ray absorption as primary cause of their soft X-ray weakness.
This makes the X-ray weak AGN important target to study the ionization structure and the matter distribution of the warm absorbers, whose presence seems to be linked to absorption in the UV band \citep{brinkmann04,piconcelli04}.
In this case, a sample composed by soft X-ray weak AGN would have a high incidence of X-ray obscured sources, playing an important role \eg\ in the X-ray background synthesis models as well as in unified AGN schemes.

On the other hand, \citet{risaliti03} from a ``mini-survey'' of $18$ quasars observed with \chandra\ found a high incidence of intrinsically under-luminous objects, suggesting that they represent a population with a Spectral Energy Distribution (SED) different from that of standard blue quasars.
So, a different intrinsic X-ray emission mechanism cannot be completely ruled out.
A strong variability of the intrinsic continuum might be another reason for a X-ray weakness at the time of the observation \citep[see \eg\ the case of \object{PG~0844+349};][]{brinkmann03}.
Recently, \citet{sch07} explained the weak X-ray emission of \object{PG~2112+059} with light bending \citep{miniutti04} instead of absorption, suggesting these sources as ideal target to observe the accretion disk near the central black hole.

In this paper we present new \xmm\ data of a soft X-ray weak QSO, \sorg, comparing the results with previous \xmm\ observations.
In our analysis, we assume a cosmology with \pedix{\Omega}{M}$=0.3$, \pedix{\Omega}{$\Lambda$}$=0.7$, and \pedix{H}{0}=$70\;$km~s$^{-1}$~Mpc$^{-1}$.


\section{\sorg}\label{sect:sorg}

\sorg\ ($z=\redgal$, \citealt{red}) is a radio-quiet \citep[ratio of radio--to--optical flux density\footnote{Measured at $6\;$cm and $4400\;$\AA, respectively.} $R=0.14$;][]{kellermann89} type~I AGN comprised in the Palomar-Green Bright Quasar Survey \citep[PG~BQS;][$\pedix{M}{B}=-20.8$ for the cosmology adopted in the present paper]{schmidt83}.
Because of the narrow \hb\ line \citep[$\mbox{FWHM(\hb)}=1480\;$km~s$^{-1}$,][]{boroson92,veron01}, the source is classified as Narrow Line Seyfert~1 (NLS1, which are historically defined to have $\mbox{FWHM(\hb)}<2000\;$km~s$^{-1}$, \citealt{osterbrock87}, and ratio of $\OIII/\hb<3$, \citealt{shuder81}).
Its optical spectrum presents strong \ion{Fe}{ii} emission lines \citep{deveny69,phillips78,smith97}, as typically observed in NLS1 galaxies \citep{boroson92,sulentic00,grupe04}.
The optical/UV continuum appears substantially reddened \citep{smith97} and polarized: its optical polarization, $P=2.5$\%, is the highest among the PG sample \citep[showing a $\langle P \rangle = 0.5$\%;][]{berriman90}.
Strong intrinsic \ion{C}{iv} $\lambda1549$ absorption features, blueshifted by $\sim 2500\;$km~s$^{-1}$ with respect to the redshift of the source \citep{sulentic06} further classify the source as a ``mini-Broad Absorption Line (BAL) QSO''\footnote{Defined as QSO showing in their optical/UV spectrum absorption lines with intermediate widths between the ``classical'' BAL QSO (absorption line width$\:\geq 2000\;$km~s$^{-1}$) and the Narrow Absorption Line QSO (absorption line width$\:\leq 500\;$km~s$^{-1}$); see \eg\ \citet{weymann91,narayanan04}.}, and imply the presence of a moderate velocity outflow.
Finally, no optical micro variability has been observed \citep{carini07}.

The mass of the central black hole of \sorg\ has been derived by \citet{vestergaard06} and \citet{zhang06} using two slightly different versions of the relation with the optical continuum luminosity and the $\mbox{FWHM(\hb)}$, and by \citet{zhang06} by rescaling the stellar velocity dispersion estimated from the \OIII\ line.
The three evaluations give $\log (\pedix{M}{BH}/\pedix{M}{$\odot$})=7.19$, $6.94$ and $7.34$, respectively.
The slightly different \pedix{M}{BH} derived from the $\mbox{FWHM(\hb)}$ and using the \OIII\ line width could suggests that the stellar velocity dispersion is larger than expected from the  $\mbox{FWHM(\hb)}-\pedix{M}{BH}$ relation.
This would imply for \sorg\ a location below the $\pedix{M}{BH}-\sigma$ relation found for local galaxies \citep[\eg,][]{tremaine02}.
A similar result was found for other NLS1 galaxies accreting close to the Eddington limit \citep{grupe04nlsy1,mathur05}, as in the case of \sorg: the bolometric luminosity and the black hole mass estimated using the \OIII\ line from \citet{zhang06} lead to an accretion rate relative to the Eddington rate of $\log \dot m = -0.32$.
It is worth noting that both NLS1s and BAL QSOs are found to lie at the extreme of the anti-correlation between \OIII\ and \ion{Fe}{ii} emission \citep[the so-called Eigenvector 1;][]{boroson92} observed in low-redshift QSO, that is thought to be driven by a high Eddington ratio \citep{boroson02}.

%
\begin{table*}
\begin{minipage}[!ht]{2\columnwidth}
\renewcommand{\thefootnote}{\thempfootnote}
\caption{EPIC {\xmm} data observation log.}             
\label{tab:xmmlog}      
\centering          
\renewcommand{\footnoterule}{}  
{\tiny
\begin{tabular}{c c c r@{\extracolsep{0.1cm}} c@{\extracolsep{0.1cm}} l l c c c c c c}
\hline\hline       
   Obs. ID    & Instrument & Filter & \multicolumn{3}{c}{Date} & Start & Exp.   & Net Exp. Time  & Net Count Rates\footnote{In the energy range $0.4 - 10\:\,$keV ({\it first line}) and $6.5 - 10\:\,$keV ({\it second line}).} & $S/N$\footnotemark[\value{mpfootnote}] & & Name used \\
              &            &        &   \multicolumn{3}{c}{}   &  End  & [ksec] & [ksec]         & [$10^{-1}\,$counts~sec$^{-1}$] & & & in the text \\
   \hline                  
                & MOS1 & thin  &    2006&Jan&24$-$25   & 22$:$16$:$06 & $25.7$ & $20.64$ &  $9.5\pm0.2$ & 43.94 & \multicolumn{1}{|c}{} & \\
                &      &       & \multicolumn{3}{c}{}  & 05$:$26$:$53 &        &         &  $0.5\pm0.1$ & 10.20 & \multicolumn{1}{|c}{} & \\
   $0300310501$ & MOS2 & thick &    2006&Jan&24$-$25   & 22$:$16$:$06 & $25.7$ & $21.60$ &  $8.6\pm0.2$ & 42.68 & \multicolumn{1}{|c}{} & \\
                &      &       & \multicolumn{3}{c}{}  & 05$:$26$:$58 &        &         &  $0.5\pm0.1$ &  9.87 & \multicolumn{1}{|c}{} & \\
                & pn   & thin  &    2006&Jan&24$-$25   & 22$:$38$:$48 & $24.0$ & $15.61$ & $34.1\pm0.5$ & 72.01 & \multicolumn{1}{|c}{} & \\
                &      &       & \multicolumn{3}{c}{}  & 05$:$27$:$13 &        &         &  $2.7\pm0.1$ & 20.04 & \multicolumn{1}{|c}{} &  {\it (A)} \\
   \cline{1-11}                    
                & MOS1 & thin  &    2006&Jan&22$-$23   & 22$:$43$:$11 & $29.3$ & $18.31$ &  $9.9\pm0.2$ & 42.11 & \multicolumn{1}{|c}{} & \\
                &      &       & \multicolumn{3}{c}{}  & 06$:$53$:$58 &        &         &  $0.5\pm0.1$ &  9.88 & \multicolumn{1}{|c}{} & \\
   $0300310401$ & MOS2 & thick &    2006&Jan&22$-$23   & 22$:$43$:$11 & $29.3$ & $18.42$ &  $8.4\pm0.2$ & 39.04 & \multicolumn{1}{|c}{} & \\
                &      &       & \multicolumn{3}{c}{}  & 06$:$54$:$03 &        &         &  $0.5\pm0.1$ &  3.91 & \multicolumn{1}{|c}{} & \\
                & pn   & thin  &    2006&Jan&22$-$23   & 23$:$05$:$53 & $27.6$ & $13.77$ & $30.4\pm0.5$ & 63.82 & \multicolumn{1}{|c}{} & \\
                &      &       & \multicolumn{3}{c}{}  & 06$:$54$:$18 &        &         &  $2.7\pm0.1$ &  5.60 & \multicolumn{1}{|c}{} & \\
   \hline
                & MOS1 & thin  &    2006&Jan&16$-$17   & 22$:$36$:$41 & $31.1$ &  $8.38$ & $14.1\pm0.4$ & 30.15 & &	\\
                &      &       & \multicolumn{3}{c}{}  & 07$:$17$:$28 &        &         &  $0.4\pm0.1$ &  5.56 & &	\\
   $0300310301$ & MOS2 & thick &    2006&Jan&16$-$17   & 22$:$36$:$41 & $31.1$ & $12.93$ & $12.3\pm0.3$ & 39.54 & & {\it (B)} \\
                &      &       & \multicolumn{3}{c}{}  & 07$:$17$:$33 &        &         &  $0.6\pm0.1$ &  8.84 & &	\\
                & pn   & thin  &    2006&Jan&16$-$17   & 22$:$59$:$23 & $29.4$ &  $5.80$ & $46.7\pm0.9$ & 51.79 & &	\\
                &      &       & \multicolumn{3}{c}{}  & 07$:$17$:$48 &        &         &  $3.2\pm0.2$ & 13.48 & &     \\
   \hline
                & MOS1 & thin  &    2002&Nov&03        & 01$:$06$:$17 & $29.5$ & $24.53$ &  $4.0\pm0.1$ & 30.71 & &	   \\
                &      &       & \multicolumn{3}{c}{} & 09$:$20$:$24  &        &         &  $0.3\pm0.1$ &  7.98 & &	   \\
   $0150610301$ & MOS2 & thin  &    2002&Nov&03        & 01$:$06$:$17 & $29.5$ & $24.53$ &  $4.1\pm0.1$ & 30.76 & & {\it Nov~2002} \\
                &      &       & \multicolumn{3}{c}{} & 09$:$20$:$24  &        &         &  $0.3\pm0.1$ &  7.87 & &	   \\
                & pn   & thin  &    2002&Nov&03        & 01$:$28$:$48 & $27.8$ & $20.66$ & $13.6\pm0.3$ & 49.58 & &	    \\
                &      &       & \multicolumn{3}{c}{} & 09$:$20$:$44  &        &         &  $1.8\pm0.1$ & 18.16 & &	    \\
\end{tabular}
}
\renewcommand{\thefootnote}{\arabic{footnote}}
\end{minipage}
\end{table*}

In the X-ray band, \rosat\ PSPC observations performed on 1993 detected the source with a $0.1-0.4\;$keV count rate\footnote{ftp://ftp.xray.mpe.mpg.de/rosat/catalogues/1rxp/} of $(0.761\pm0.195)\times10^{-2}\;$cts~s$^{-1}$.
The source was not detected in the hard ($0.5-2.0\;$keV) band, nor it is detected in the \rosat\ All Sky Survey catalog\footnote{http://www.xray.mpe.mpg.de/cgi-bin/rosat/src-browser}. 
\citet{gallagher01} estimated an upper limit to the observed count rate in the $0.5-2\;$keV energy range\footnote{The authors reported the detection of $5$ photons above $0.5\;$keV in a $1.\arcmin 2$ radius source cell, with $2$ expected from the background.} of $3.2\times10^{-3}\;$cts~s$^{-1}$.
The non-detection above $0.5\;$keV reminds another NLS1, the X-ray transient \object{WPVS~007}.
In their analysis of \rosat\ data, \citet{grupe95} suggested obscuration as the possible cause of the turn-off observed in the $0.5-2.4\;$keV energy range.
On the basis of the \rosat\ observation, \citet{brandt00} classified \sorg\ as soft X-ray weak QSO.
In 1999, six years after the \rosat\ observations, \asca\ found that the source emission has varied in a significant way: assuming the best fitting \asca\ spectral model, \citet{gallagher01} predicted a PSPC count rate of $4.9\errUD{1.1}{0.9}\times10^{-3}\;$cts~s$^{-1}$, \ie\ a factor $\sim 1.5$ higher than the upper limit estimated for the \rosat\ observation.
Moreover, the authors found evidence of X-ray intrinsic absorption ($\nhsym\approx10^{23}\;$\nh) with only partial covering of the power law continuum.
The limited quality of the \rosat\ data prevented the authors from ascribing the observed variation to an increase of the continuum flux or a decrease of the absorption.
As noted by \citet{gallo06}, NLS1 galaxies with complex high-energy spectra (such as \sorg) appear to be X-ray weak AGN.

The \xmm\ observation performed in \dataO\ found \sorg\ in a higher state than previous observations at energies $E>2\;$keV \citep{sch05}.
Moreover, the EPIC spectra clearly require the presence of complex absorbing structure.
A good description of the data is obtained only assuming a combination of an ionized absorber and a neutral absorber with a large covering factor.
Finally, the data suggest the presence of a broad emission feature, whose profile is well described by a relativistically broadened \feka\ emitted in a disk accreting around a black hole.
The absence of a features like this in the \asca\ data allowed the authors to suggest a possible variability in the emission line.
A similar behaviour (\ie, complex X-ray emission and strong variability) has been reported for other NLS1s: for example, \citet{grupe07} found \object{Mrk~335} in an extremely low X-ray state, whose interpretation includes partial-covering absorption or X-ray reflection from the disk.
A comparison of the emission observed in 2007 with all previous observations shows a decrease in flux by a factor of more than $30$, that the authors ascribe mainly to a change of the intrinsic absorption.


\section{Observation and data reduction}

We observed \sorg\ (\posgal) with \xmm\ in \dataN\ in three pointings, the first separated by about one week from the others, for a total of about $90\:\,$ksec (Obs.~ID $0300310301$, $0300310401$ and $0300310501$). 
The observations were performed with the European Photon Imaging Camera (EPIC), the Optical Monitor (OM) and the Reflection Grating Spectrometer \citep[RGS;][]{rgs}; the source is not detected with the last instrument.
The three EPIC cameras \citep[pn, MOS1, and MOS2;][]{pn,mos} were operating in full frame mode, MOS1 and pn with the thin filter and MOS2 with the thick filter applied.
The observation details are reported in Table~\ref{tab:xmmlog}.
OM \citep{om} performed 6 observations for each pointing, three with the UVM2 ($\lambda_{eff}=2310\;$\AA) and three with the UVW2 ($\lambda_{eff}=2120\;$\AA) filters in the optical light path, of $1.9\;$ksec each; all OM exposures were performed in the ``Science User Defined'' image mode (with a windows size of $5\arcmin \times 5\arcmin$). 
The \xmm\ data have been processed using the Science Analysis Software (SAS version~7.0) with the calibration from February~2007; the tasks \emph{epproc} and \emph{emproc} were run to produce calibrated and concatenated event lists for the EPIC cameras.

EPIC event files have been filtered for high-background time intervals, following the standard method consisting in rejecting periods of high count rate at energies $>10\:\,$keV.
Events corresponding to patterns $0 - 12$  (MOS1\&2) and $0 - 4$ (pn) have been used \citep[see the \xmm\ Users' Handbook;][]{xmmhb}.
We have also generated the spectral response matrices at the source position using the SAS tasks \emph{arfgen} and \emph{rmfgen}.
We can exclude event pile-up in the MOS and pn data.
Source counts were extracted from a circular region of radius $18\arcsec$ and $21\arcsec$ for the MOS and pn.
Background counts were extracted from a nearby source-free circular region of $36\arcsec$ and $42\arcsec$ radii, respectively. 
The net exposure times at the source position, the net count rates and the $S/N$ ratios in the energy ranges $0.4-10\;$keV and $6.5-10\;$keV are reported in Table~\ref{tab:xmmlog}.

We tested short time variability within each observation, generating source light curves in different energy ranges with a binning time of $5\;$ksec.
No bin shows significant deviation from the mean value.

We processed OM data with the \emph{omichain} routine of SAS; for each OM filter and for each pointing three images were obtained. 
The count rate informations were extracted from the three \emph{omichain} output and averaged, assuming the standard deviation as an estimate of their error.
The count rates were converted to magnitudes (in the AB system) and fluxes using the conversion factors reported by the SAS homepage\footnote{http://xmm.vilspa.esa.es/sas/7.0.0/watchout/\\Evergreen\_tips\_and\_tricks/uvflux.shtml}; the errors have been calculated propagating the uncertainties on the count rates.
The obtained values are provided in Table~\ref{tab:omfl}; in all the observations, the ratio of the maximum to the minimum count rates is consistent with unity within the errors.

In order to better compare the old and new \xmm\ observations, we have re-analysed the EPIC \xmm\ data taken in \dataO\ (archived under the observation identifier $0150610301$).
At this occasion, \sorg\ has not been observed with OM.
Following \citet{sch05}, for the two MOS exposures the background was determined from an annulus around the extraction area of the source counts (inner and outer radii: $48\arcsec$ and $84\arcsec$, respectively). 
Details of this observation, as well as exposure times, count rates and $S/N$ ratios in the energy ranges $0.4-10\;$keV and $6.5-10\;$keV, are reported in Table~\ref{tab:xmmlog}.

%
\begin{table}[t]
\begin{minipage}[h]{\columnwidth}
\caption{OM integrated magnitudes and fluxes (\dataN\ observations).}             
\label{tab:omfl}      
\centering          
{\tiny
\begin{tabular}{c@{\extracolsep{0.2cm}} c@{\extracolsep{0.10cm}} c@{\extracolsep{0.2cm}} c@{\extracolsep{0.10cm}} c}
\hline\hline       
   Obs. ID    & \multicolumn{2}{c}{AB magnitudes} & \multicolumn{2}{c}{Flux}\\
              &  \multicolumn{2}{c}{[mag]}        & \multicolumn{2}{c}{[$10^{-15}\,$\fluxA]}\\
  \cline{2-3} \cline{4-5}
              & UVM2 & UVW2 & UVM2 & UVW2 \\
              & [$2310\;$\AA] & [$2120\;$\AA] & [$2310\;$\AA] & [$2120\;$\AA] \\
\hline                    
   $0300310501$ & $16.567\pm0.016$ & $16.719\pm0.027$ & $4.811\pm0.069$ & $4.977\pm0.123$ \\
   $0300310401$ & $16.510\pm0.015$ & $16.648\pm0.026$ & $5.072\pm0.071$ & $5.312\pm0.127$ \\
   $0300310301$ & $16.496\pm0.015$ & $16.625\pm0.025$ & $5.139\pm0.071$ & $5.427\pm0.127$ \\
\hline                  
\end{tabular}
}
\end{minipage}
\end{table}
%


\section{The X-ray emission: a first comparison between the datasets}

We performed the spectral analysis of the MOS and pn spectra over the energy range from $0.4\;$keV to $10.0\;$keV.
The EPIC spectra have been analysed using standard software packages \citep[FTOOLS version~6.2, XSPEC version~12.3.1q, XSPEC version~11.3.2ag;][]{xspec}.
All the models discussed in the following assume Galactic absorption with a column density of $\pedix{N}{H, Gal}=\nhgal\:\,$\nh\ \citep{nh}.
Unless otherwise stated, the figures are in the rest-frame of \sorg, and fit parameters are quoted in the same frame.

Comparing the three observations of \dataN\ it is evident that the flux decreases considerably passing from the observation $0300310301$ to the other two, that present the same spectral shape and intensity.
This behaviour is particularly enhanced at energies lower than $0.7\;$keV.
Therefore we combined the pn and MOS data from $0300310401$ and $0300310501$ (the label A identifies the combined spectra), keeping separate the $0300310301$ data, that are labeled as B.
In order to apply the $\chi^2$ statistics, source counts for the (A) and (B) spectra were binned to have at least $40$ and $30$ counts in each energy bin, respectively.

\begin{figure}[t]
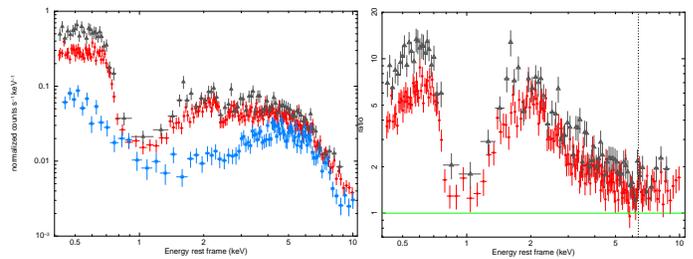

  \centering
  \resizebox{\hsize}{!}{\includegraphics[angle=270]{f1a.ps}\includegraphics[angle=270]{f1b.ps}}
  \caption{{\it Left panel:} EPIC-pn spectra of the three observations: red filled circles, spectrum (A); grey open triangles, spectrum (B); sky-blue stars, \dataO\ data.              {\it Right panel:} Ratios between the new EPIC-pn spectra and the best fit model for the \dataO\ dataset \citep{sch05}.
      	   The black dotted line corresponds to the position of the $6.4\;$keV line.}
  \label{fig:cfrobs}%
\end{figure}

From a comparison between the new and the old datasets, a strong change of both flux and spectral shape is evident (see Fig.~\ref{fig:cfrobs}, {\it left panel}).

In Fig.~\ref{fig:cfrobs}, {\it right panel}, we have plotted the ratio of the \dataN\ \xmm\ spectra of \sorg\ to the \dataO\ data (considering that the MOS spectra agree, within the errors, with the EPIC-pn spectra, to avoid clutter only the last are shown).
The best fit found by \citet{sch05} for the \dataO\ data (an intrinsic power law with superimposed a relativistically broadened line emitted by an accretion disk around a Kerr black hole, covered by a combination of an ionized absorber and a neutral absorber with a large covering factor) cannot describe neither the (A) data or the (B) spectra.
This figure may offer a first indication about the origin of the observed emission.
The strongest variation is observed in the soft energy band, with the emission between $0.7$ and $1.5\;$keV being less affected by changes.
This ratio would imply that the observed variability is associated mainly with absorbing matter, whose physical properties are different among the observations.

%
%


\section{Spectral analysis of the \dataN\ (A) data}

We started with the (A) data, due to the higher $S/N$ of these spectra.
The analysis of the \dataO\ and the (B) datasets was performed taking into account the results of this first part of the work.

\subsection{The emission at energies $>3.5\;$keV}\label{sect:highA}

%
\begin{table*}
\begin{minipage}[t]{2\columnwidth}
\caption{\footnotesize Analysis of the high energy ($3.5-10\;$keV) part of the \dataN\ (A) pn and MOS spectra.}
\label{tab:obs23high}
\begin{center}
\renewcommand{\footnoterule}{}  
{
\begin{tabular}{c@{\extracolsep{0.8cm}} r@{\extracolsep{0.18cm}} c@{\extracolsep{0.13cm}} c@{\extracolsep{0.13cm}} c@{\extracolsep{0.18cm}} c@{\extracolsep{0.13cm}} c@{\extracolsep{0.13cm}} c@{\extracolsep{0.18cm}} c}
 \multicolumn{9}{l}{Results for a reflection-based model.} \\                 
  \cline{2-9}
   & & \multicolumn{3}{c}{Power law+Gaussian$^{\mydag}$} & \multicolumn{3}{c}{Kdblur[Reflion]$^{\myddag}$}  \\
  \cline{3-5} \cline{6-8}
   &
  \multicolumn{1}{c}{Model} 
   & $\Gamma$
   & Norm$^a$
   & $\pedSM{E}{rf}\,$[keV] 
   & $\beta$
   & $\pedSM{R}{in}\,[GM/c^2]$
   & $i\,[\arcdeg]$
   & \chidof \\
   &  
   &
   &  
   &  
   & Fe/solar
   & $\xi$$^b$
   & Norm$^c$
   & \\
  \cline{2-9}
   &  {disk$+$line} & $2.32\errUD{0.96}{0.13}$ & $6.40\errUD{1.89}{1.28}$ & $6.46\pm0.04$ & $3.02\errUD{0.27}{0.39}$ & $3.85\errUD{1.16}{2.39}$ & $0.02\errUD{0.05}{0.01}$ & $109.2/119$ \\
   &   &  &  &  & $1.25\errUD{4.99}{0.22}$ & $159.59\errUD{177.14}{127.41}$ & $4.26\errUD{0.24}{0.17}$ & \\
  \multicolumn{9}{l}{} \\                 
  \multicolumn{9}{l}{Results for an absorption-based model.} \\                 
  \cline{2-9}     
  & & \multicolumn{3}{c}{Power law+Gaussian$^{\mydag}$} & \multicolumn{2}{c}{PC absorber} & Absorber & \\
  \cline{3-5} \cline{6-7} \cline{8-8}
   &
  \multicolumn{1}{c}{Model} 
   & $\Gamma$
   & Norm$^a$
   & $\pedSM{E}{rf}\,$[keV] 
   & \nhsym$^d$
   & Cov.~fraction$^e$ 
   & \nhsym$^d$
   & \chidof \\
  \cline{2-9}                     
   & {neutral pc$+$line} & $2.55\errUD{0.25}{0.20}$ & $40.87\errUD{10.82}{13.34}$ & $6.41\errUD{0.07}{0.05} $ & $10.79\errUD{1.62}{2.22}$ & $0.80\errUD{0.20}{0.09}$ & $3.24\errUD{1.08}{1.11}$ & $121.3/123$ \\
\end{tabular}
}
\end{center}       
{\scriptsize  Errors are quoted at the 90\% confidence level for 1 parameter of interest ($\Delta\apix{\chi}{2}=2.71$). $^{\mydag}\;$Narrow line width fixed to $0\;$eV. $^{\myddag}\;$Photon index coupled with the primary power law photon index; outer radius fixed to the maximum value, $\pedSM{R}{out}=400\,$\pedSM{r}{g}. $^a\;$In units of $10^{-4}\,$\normPL\,@$1\,$keV. $^b\;$Ionization parameter $\xi \equiv 4\pi \pedSM{F}{ill}/n$ (erg~cm~s$^{-1}$), where \pedSM{F}{ill} is the ionizing flux and $n$ is the hydrogen nucleus density (part~cm$^{-3}$) of the illuminated slab. $^c\;$In units of $10^{-6}\,$\normPL\,@$1\,$keV of the reflected spectrum. $^d\;$In units of $10^{22}\,$\nh. $^e\;$Complement to one of the covered-to-uncovered flux ratio. } \\
\end{minipage}
\end{table*}

\begin{figure}[t]
  \centering
  \resizebox{\hsize}{!}{\includegraphics[angle=270]{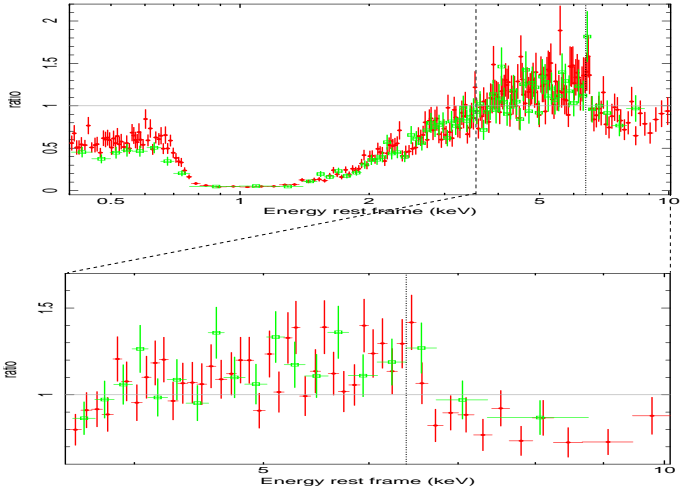}}
  \caption{Data/model ratio for the (A) spectra (green open squares, MOS; red filled circles, pn).
           The adopted model is a simple power law with Galactic absorption fitted between $3.5$ and $5\;$keV, and extrapolated up to $10\;$keV ({\it lower panel}), and to the whole studied energy range ($0.4-10\;$keV, {\it upper panel}; for demonstration purposes, here the data have been binned to have a significance of $8\sigma$ up to 30 bins).
	   In both panels, the black dotted line marks the position of the $6.4\;$keV line.}
  \label{fig:obs23pl}%
\end{figure}

First, we fitted the (A) data with a simple power law.
Due to the complex spectral shape both at low and high energies, we considered only the $3.5-5\;$keV energy range, finding a photon index $\Gamma = 1.40 \pm 0.19$.
This spectral index appears flatter than the mean value found at energies $>2\;$keV for a sample of AGN selected in the PG~BQS \citep[$\langle \Gamma \rangle = 1.87 \pm 0.10$;][]{piconcelli05}.
This discrepancy is even stronger considering only NLS1 galaxies, whose photon indices appear to be on average steeper than those of Seyfert galaxies with broad optical lines (\eg, studying \asca\ spectra of $23$ NLS1s, \citealt{leighly99} found a mean value in the $2-10\;$keV energy range $\langle \Gamma \rangle = 2.19 \pm 0.10$).
When the model is extrapolated to the whole $0.4-10\;$keV band, the most evident feature is a strong decrease in the flux below $3\;$keV, particularly enhanced between $0.7$ and $2\;$keV (Fig.~\ref{fig:obs23pl}, {\it upper panel}).

Focusing at $E>3.5\;$keV, this simple model shows residual structures in the $5.5-6.5\;$keV rest frame region, and is also unable to describe the high energy end of the spectra (Fig.~\ref{fig:obs23pl}, {\it lower panel}).

%

To model the observed residuals, as a first step we added one or more narrow or broad line(s) to the powerl-law continuum (ZGAUSS component in XSPEC, with $\sigma=1\;$eV or free to vary), and/or an absorption edge at $\sim 7\;$keV (ZEDGE component in XSPEC).
The addition of the absorption edge can't reproduce the data between $6$ and $7\;$keV, and still required an unusually low photon index, $\Gamma\sim 1.27$.
The observed spectra can be well described assuming a model with two Gaussian lines, one narrow and one broad, whose energies were consistent with being produced by neutral iron, or a power law model with superimposed a disk line in a Kerr configuration \citep[LAOR component in XSPEC;][]{laor91,fabian02}.
The addition of a narrow Gaussian line with energy consistent with being produced by matter in low ionization stages (\ie, \ion{Fe}{i}-\ion{Fe}{xvii}) improves the fit significantly ($F$-test probability\footnote{But see the caveats in using the F-test to measure the significance of narrow lines described in \citet{protassov02}.} $97.3$\%).
Being the line unresolved by EPIC-pn, in the following its width has been fixed to the instrumental spectral resolution.

In the last years, two different scenarios have been proposed to account for high-energy spectral shapes as observed in the (A) spectra (Fig.~\ref{fig:obs23pl}, {\it lower panel}): a ``reflection-based'' (\eg, \object{Mrk~335}, \citealt{longinotti07,oneill07}; \object{HE~0450-2958}, \citealt{zhou07}) or an ``absorption-based'' (\eg, \object{1H~0707-495}, \citealt{boller02,gallo04,fabian04,tanaka04}; \object{1H~0419-577}, \citealt{pounds04,fabian05}; \object{IRAS~13197-1627}, \citealt{dadina04,miniutti07}) one.
Despite their different shape in the Fe line energy region, the superposition of different layers of partially ionised absorbing material could mimic the presence of a broad line component superimposed to the continuum, a feature naturally present in relativistically blurred ionized reflection model \citep[see \eg][]{reeves04,grupe07,petrucci07}.

A reflection component can reproduce in a good way ($\chidof =109.2/119$) the high-energy curvature only if originates in the accretion disk\footnote{Reflection from neutral material far from the nucleus \citep[PEXRAV in XSPEC;][]{pexrav} does not provide a good description of the data ($\chidof =132.2/121$).}, leading to a rather steep intrinsic continuum, $\Gamma\sim 2.3$.
To model the ionized reflection in XSPEC we used the REFLION\footnote{http://heasarc.gsfc.nasa.gov/docs/xanadu/xspec/models/reflion.html} external table, which incorporates both line emission with Compton broadening and reflection continuum \citep{ross05,ross99}.
Then, the reflection from the disk has been constructed convolving this component with the relativistic blurring model KDBLUR, produced by A.~Fabian and R.~Johnstone using the same kernel of the LAOR disk line model\footnote{Unless otherwise stated, the outer radius was fixed to its maximum value, $\pedix{R}{out}=400\:\pedix{r}{g}$, while the inner radius and the emissivity power law index were free to vary.}.

The $3.5 - 10\;$keV data can be equally well reproduced ($\chidof =121.3/123$) in an absorption scenario, assuming a high column density ($\nhsym \sim10^{23}\;$\nh) cold absorber covering $\sim 80$\% of the primary continuum. 
This latter component is rather steep ($\Gamma \sim 2.5$), to account for the observed spectral drop at the highest energies ($E>7\;$keV). 
Allowing the ionization state of the absorbers or the iron abundance to be free parameters, the fit does not improve further. 

\begin{figure}[t]
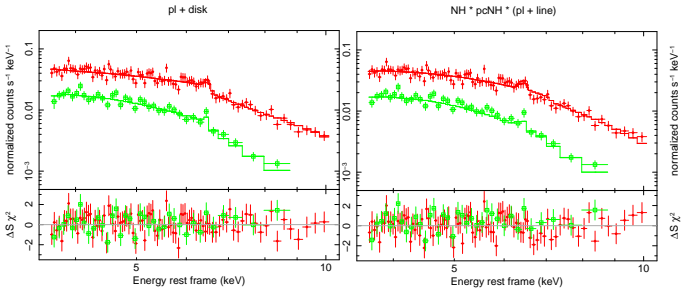

  \centering
  \resizebox{\hsize}{!}{\includegraphics[angle=270]{f3a.ps}
  \includegraphics[angle=270]{f3b.ps}}  
  \caption{EPIC-pn \dataN\ (A) spectra of \sorg\ in the $3.5-10\;$keV energy range, plotted in comparison to the reflection-based ({\it left panel}) and the absorption-based  ({\it right panel}) models (see \S\ref{sect:highA} and Table~\ref{tab:obs23high}).}
  \label{fig:high23}%
\end{figure}

The high-energy curvature observed in the (A) spectra of \sorg\ can thus be well accounted for in both reflection-based and absorption-based scenarios (see Table~\ref{tab:obs23high} and Fig.~\ref{fig:high23}).
A gaussian emission line is detected with an $F$-test probability of $98.6$\% in both cases; its energy and EW ($\sim 50-80\;$eV) are compatible with K$\alpha$ fluorescence emission of almost neutral iron, produced by transmission through cold gas of column density $<10^{23}\;$\nh\ totally covering the source \citep{makishima86}.

%

%
\begin{table*}
\begin{minipage}[!h]{2\columnwidth}
\caption{\footnotesize Analysis of the broad band \dataN\ (A) pn and MOS spectra ($0.4-10\;$keV).}
\label{tab:broad}
\begin{center}   
\renewcommand{\footnoterule}{}  
{\tiny
\begin{tabular}{c@{\extracolsep{0.8cm}} r@{\extracolsep{0.18cm}} c@{\extracolsep{0.13cm}} c@{\extracolsep{0.18cm}} c@{\extracolsep{0.13cm}} c@{\extracolsep{0.18cm}} c@{\extracolsep{0.13cm}} c@{\extracolsep{0.13cm}} c@{\extracolsep{0.18cm}} c@{\extracolsep{0.13cm}} c@{\extracolsep{0.13cm}} c@{\extracolsep{0.13cm}} c}
 \vspace{-0.2cm}  \\            
 \multicolumn{13}{l}{Result for a warm absorber$*$[partial covering warm absorber$*$(intrinsic power law$+$disk reflection)$+$narrow line] model.} \\                 
\cline{2-13}
   & & \multicolumn{2}{c}{Power law} & \multicolumn{3}{c}{PC warm absorber} & \multicolumn{2}{c}{Warm absorber} & \multicolumn{3}{c}{Kdblur[Reflion]$^{\mydag}$} &  \\
  \cline{3-4} \cline{5-7} \cline{8-9} \cline{10-12}
   & 
   Obs.
   & $\Gamma$
   & Norm$^a$
   & \nhsym$^b$
   & $\xi$$^c$
   & Cov.~fraction$^d$
   & \nhsym$^b$
   & $\xi$$^c$
   & $\beta$
   & $\pedSM{R}{in}\,$[\pedSM{r}{g}]
   & $\xi$$^c$
   & \chidof \\
   & 
   & 
   &
   &  
   &  
   &  
   & 
   & 
   & $i\,[\arcdeg]$ 
   & Fe/solar
   & Refl.~fraction$^e$
   & \\
\cline{2-13}
   &  {\it (A)} & 2.06$\pm$0.08 & 3.21\errUD{1.81}{1.03} & 6.11\errUD{2.04}{1.36} & 68.14\errUD{34.15}{20.77} & 0.78\errUD{0.08}{0.12} & 2.13$\pm$0.08 & 3.67$\pm$0.08 & 4.34\errUD{0.34}{0.27} & 1.31\errUD{1.39}{0.07} & 121.59\errUD{40.48}{52.22} & $290.1/281$ \\
   &   &  &  &  &  &  &  &  & $0.003\pm0.01$ & 4.54\errUD{0.35}{1.01} & $65.2$\% & \\
 \multicolumn{13}{l}{} \\                 
 \multicolumn{13}{l}{Result for a neutral absorber$*$\{warm absorber$*$[partial covering warm absorber$*$(intrinsic power law)]$+$scattered component$+$narrow lines\} model.} \\
 \cline{2-13}
   & & \multicolumn{2}{c}{Power law$^{\myddag}$} & \multicolumn{3}{c}{PC warm abs.} & \multicolumn{2}{c}{Warm abs.} & \multicolumn{2}{c}{Gussian} & Neutral abs. & \\
  \cline{3-4} \cline{5-7} \cline{8-9} \cline{10-11} \cline{12-12}
   & 
   Obs.
   & $\Gamma$
   & Norm$^{a}$
   & \nhsym$^b$
   & $\xi$$^c$ 
   & Cov.~fraction$^d$
   & \nhsym$^b$
   & $\xi$$^c$ 
   & \pedSM{E}{rf}$^f$
   & $\sigma$$^g$
   & \nhsym$^b$
   & \chidof \\
   & 
   & \pedSM{\Gamma}{scatt}
   & Scatt.~fraction$^h$
   & 
   & 
   & 
   & 
   & 
   & 
   & 
   & 
   & \\
\cline{2-13}
   & {\it (A)} & $2.43\errUD{0.13}{0.03}$ & $22.74\errUD{10.19}{8.06}$ & $8.43\errUD{1.26}{1.54}$ & $19.96\errUD{4.53}{3.15}$ & $0.43\errUD{0.03}{0.11}$ & $4.49\errUD{0.17}{0.12}$ & $94.04\errUD{0.98}{1.73}$ & $0.59\errUD{0.01}{0.02}$ & $< 30$ & $0.21\errUD{0.02}{0.02}$ & $297.5/279$ \\
   &           & $2.83\errUD{0.39}{0.67}$ & $0.006\errUD{0.002}{0.001}$ &  &  &  &  &  &  &  &  &  \\
 \multicolumn{13}{l}{} \\                 
\end{tabular}
}
\end{center}
{\scriptsize  Errors are quoted at the 90\% confidence level for 1 parameter of interest ($\Delta\apix{\chi}{2}=2.71$); narrow line energy fixed to $\pedSM{E}{rf}=6.4\;$keV, narrow line width fixed to $0\;$eV. $^{\mydag}\;$Photon index coupled with the primary power law photon index; outer radius fixed to the maximum value, $\pedSM{R}{out}=400\,\pedSM{r}{g}$. $^{\myddag}\;$First line, intrinsic power law; second line, scattered power law. $^a\;$In units of $10^{-4}\,$\normPL\,@$1\,$keV. $^b\;$In units of $10^{22}\,$\nh. $^c\;$Ionization parameter $\xi \equiv 4\pi \pedSM{F}{ill}/n$ (erg~cm~s$^{-1}$), where \pedSM{F}{ill} is the ionizing flux and $n$ is the hydrogen nucleus density (part~cm$^{-3}$) of the illuminated slab. $^d\;$Complement to one of the covered-to-uncovered flux ratio. $^e\;$Ratio between the ionized reflection flux and the total flux in the $(0.1-1000)\,$keV energy range. $^f\;$Line centroid energy, in keV. $^g\;$Line width, in eV. $^h\;$Ratio between the scattered flux and the primary, unabsorbed flux.}  \\
\end{minipage}
\end{table*}

\subsection{The whole energy range}\label{sect:wholeA}

\begin{figure*}[!h]
  \centering
  \resizebox{\hsize}{!}{\includegraphics[angle=270]{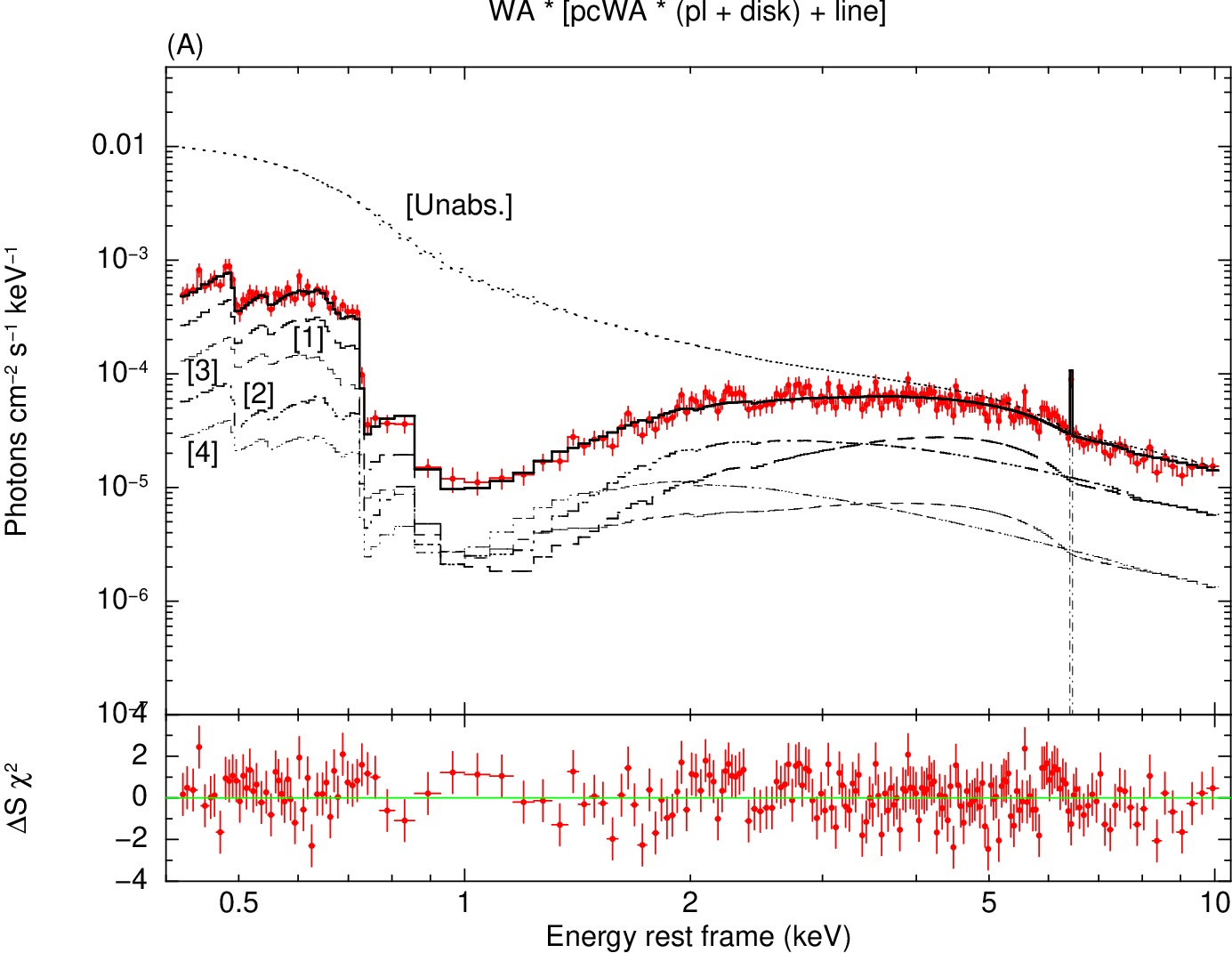}\includegraphics[angle=270]{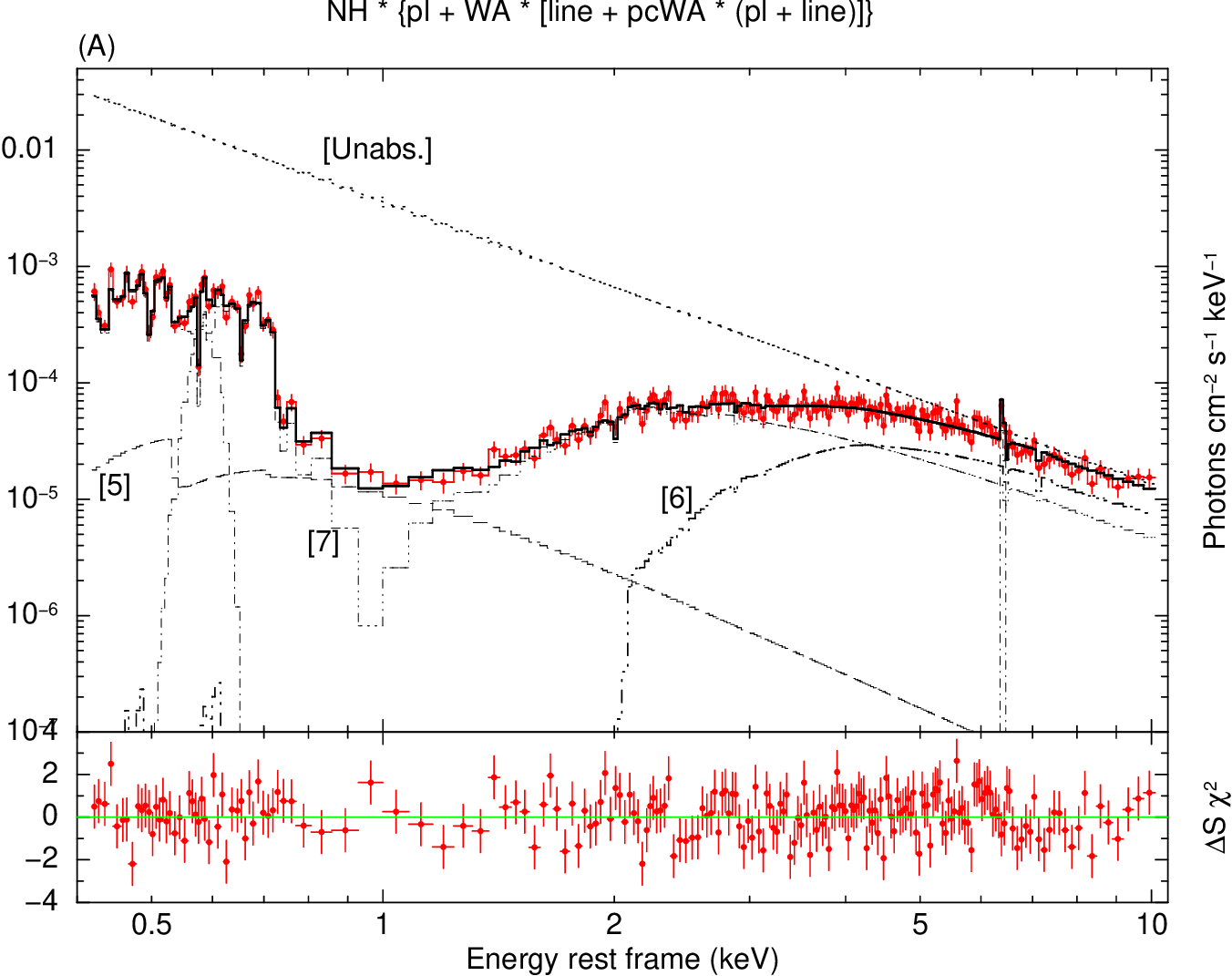}}
  \caption{Comparison of the unfolded best fits to the EPIC-pn \dataN\ (A) data of \sorg\ with the two adopted models (see \S\ref{sect:wholeA}; {\it upper panels}).
	   The residuals in terms of sigmas are plotted in the {\it lower panels}.
           The solid lines correspond to the total best fit models; over-plotted are the contributions of the different components.
	   The unabsorbed emission, labeled as [Unabs.], is also shown.
           {\it Left panels}: Reflection from the disk (dashed line) and intrinsic power law (dash-dot-dotted line) partially covered by an warm absorber (the thickest lines identify the fraction of continuum covered, [1] and [2] respectively; the thinnest lines identify the fraction uncovered, [3] and [4] respectively), with a narrow line superimposed (dash-dotted line); the whole continuum was seen through a secon ionized layer of material.
           {\it Right panels}: Scattered component (dashed line, [5]), absorbed power law (dash-dot-dotted line) and narrow emission lines (dash-dotted line), partially covered by an warm absorber (the thickest line identify the fraction of continuum covered, [6], the thinnest the fraction uncovered, [7]); the whole continuum was seen through layer of neutral material.}
  \label{fig:ufde23}%
\end{figure*}

We then extended the analysis at all energies ($0.4-10\;$keV).
The (A) spectra between $\sim 0.7$ and $2\;$keV appear strongly absorbed, with an important component rising in the softer band.
Evident absorption features are present at $E\sim2\;$keV as well as at energies lower than about $0.7\;$keV.
Simple models, composed by a single power law continuum covered by a combination of neutral or ionized absorbers \citep[modeled with ABSORI component in XSPEC;][]{absori}, were not able to reproduce the observed shape.
As obtained considering only the high-energy part of the spectra, a model based on reflection from neutral material distant from the black hole alone was not able to explain the observed emission, too.
More complex scenarios have to be investigated.

The basic models adopted to test the reflection-based and the absorption-based scenarios are the ones described in \S\ref{sect:highA}:
\begin{itemize}
 \item a power law plus reflection from the disk (the latter constructed in XSPEC using the REFLION external table convolved with the KDBLUR model);
 \item a simple power law and a complex partial covering material.
\end{itemize}
The overall broad-band shape was roughly reproduced by the two models, although in the latter case the strong soft excess observed can be partially accounted for only allowing the absorbing column densities to be ionized.
Under this assumption, the absorbers are able to let the nuclear flux to be transmitted at low energies. 
However, these simple models were inadequate to model the absorption features, resulting in statistically unacceptable fits.

\vspace{0.5cm}
For the reflection-based model, following the best fit obtained in \S\ref{sect:highA} we adopted a continuum model composed by an intrisic power law and a relativistically blurred ionized reflection.
To model the absorption features observed in the spectra, we assumed the presence of a warm absorber partially covering the central region; a second distribution of matter with  lower ionization covering the whole emission is also required.
Finally, a transmitted narrow line at $6.4\;$keV (rest frame) completes the model.
This complex model resulted to be a good description of the (A) data ($\chidof=290.1/281$); the best fit parameters are shown in the first part of Table~\ref{tab:broad}.

As previously noted, the presence of ionized absorption can affect the high-energy spectrum and the iron line profile.
We then checked our results modeling the absorber in a more sophisticated way, adopting a more refined photoionization model, namely XSTAR\footnote{http://heasarc.gsfc.nasa.gov/docs/software/xstar/xstar.html} \citep{xstar}.
XSTAR is a public code that consistently (\ie, by solving temperature and ionization balance equations) calculates the opacities of a spherical shell of Compton-thin gas photoionized by a central source of radiation. 
It offers the possibility to construct tables varying the photoionizing continuum spectral shape and strength, the gas physical conditions (density, pressure, thickness) and the abundances of cosmic elements.
Its atomic database includes all the most relevant transitions.
To perform the fit using XSPEC version~12\footnote{The model assumed to describe the reflection from the disk taking into account the relativistic blurring can be fitted only inside the new version of XSPEC. On the other hand, using the XSTAR tables in the last version of XSPEC, the minimization algorithm suffers problems in interpolating the column density values between grid points. 
Their origin is currently under investigation (Craig~Gordon, priv. comm.).}, we considered \nhsym\ to be a free parameter but keeping its value fixed during the minimization.
The best fitting parameters are in good agreement with the values reported in the first part of Table~\ref{tab:broad}, while the fit obtained with this change in the absorption model is statistically worse.
Above all, a disk component is still statistically required.

%
\begin{table*}
\begin{minipage}[!h]{2\columnwidth}
\caption{\footnotesize Simultaneous analysis of the three available \xmm\ datasets for \sorg\ in the $0.4-10\;$keV energy range: the adopted model was the accretion disk reflection model (a continuum composed by an intrinsic power law and a reflection from the disk partially covered by an ionized absorber; the whole continuum was seen through a warm absorber; $\chidof=379.6/364$).
Luminosity are absorption-corrected, while fluxes are corrected only for the Galactic absorption.}
\label{tab:allbfDisk}
\begin{center}   
\renewcommand{\footnoterule}{}  
{\tiny
\begin{tabular}{r@{\extracolsep{0.18cm}} c@{\extracolsep{0.13cm}} c@{\extracolsep{0.18cm}} c@{\extracolsep{0.13cm}} c@{\extracolsep{0.18cm}} c@{\extracolsep{0.13cm}} c@{\extracolsep{0.13cm}} c@{\extracolsep{0.18cm}} c@{\extracolsep{0.13cm}} c@{\extracolsep{0.13cm}} c}
 \vspace{-0.2cm}  \\            
  \hline\hline  
   & \multicolumn{2}{c}{Power law} & \multicolumn{3}{c}{PC warm absorber} & \multicolumn{2}{c}{Warm absorber} & \multicolumn{3}{c}{Kdblur[Reflion]$^{\mydag}$} \\
  \cline{2-3} \cline{4-6} \cline{7-8} \cline{9-11}
   Obs.
   & $\Gamma$
   & Norm$^a$
   & \nhsym$^b$
   & $\xi$$^c$
   & Cov.~fraction$^d$
   & \nhsym$^b$
   & $\xi$$^c$
   & $\beta$
   & $\pedSM{R}{in}\,$[\pedSM{r}{g}]
   & $\xi$$^c$ \\
   & 
   &
   &  
   &  
   &  
   & 
   & 
   & $i\,[\arcdeg]$ 
   & Fe/solar
   & Refl.~fraction$^e$ \\
  \hline                    
   {\it (A)} & $2.05\errUD{0.05}{0.12}$ & $2.76\errUD{0.98}{0.77}$ & $5.77\errUD{1.97}{1.90}$ & $55.19\errUD{25.86}{24.74}$ & $0.72\pm0.08$ & $2.07\errUD{0.18}{0.28}$ & $3.28\errUD{0.70}{1.46}$ & $4.26\errUD{0.23}{0.27}$ & $1.31\errUD{1.20}{0.07}$ & $130.71\errUD{10.11}{19.31}$ \\
    &  &  &  &  &  &  &  & $(2.7\pm0.6)\times10^{-3}$ & $4.71\errUD{0.87}{1.83}$ & $48.8$\% \\
   {\it (B)} & $2.05^{\myddag}$ & $2.57\errUD{0.55}{1.01}$ & $3.44\errUD{2.13}{1.28}$ & $66.62\errUD{39.88}{32.57}$ & $0.85\pm0.06$ & $2.07^{\myddag}$ & $3.28^{\myddag}$ & $4.26^{\myddag}$ & $1.31^{\myddag}$ & $125.97\errUD{12.14}{9.65}$ \\
    &  &  &  &  &  &  &  & $(2.7\times10^{-3})^{\myddag}$ & $4.71^{\myddag}$ & $60.1$\% \\
   {\it Nov.~2002} & $1.79\errUD{0.10}{0.09}$ & $1.55\errUD{0.35}{0.96}$ & $20.94\errUD{4.46}{2.87}$ & $98.87\errUD{30.08}{25.81}$ & $0.92\errUD{0.03}{0.01}$ & $0.54\errUD{0.25}{0.36}$ & $3.28^{\myddag}$ & $4.26^{\myddag}$ & $1.31^{\myddag}$ & $65.43\errUD{21.50}{17.23}$ \\
    &  &  &  &  &  &  &  & $(2.7\times10^{-3})^{\myddag}$ & $4.71^{\myddag}$ & $45.0$\% \\
  \hline
  \multicolumn{11}{l}{} \\                 
    & \multicolumn{4}{c}{Observed flux $[10^{-13}\,\flux]$} & \multicolumn{4}{c}{Unabsorbed luminosity $[10^{43}\,\lum]$} \\     
    \cline{2-5}\cline{6-9}          
   Obs.
    & $0.4-10\;$keV
    & $0.4-1\;$keV
    & $1-3\;$keV
    & $3-10\;$keV 
    & $0.4-10\;$keV
    & $0.4-1\;$keV
    & $1-3\;$keV
    & $3-10\;$keV \\
    \cline{1-9}          
    {\it (A)} & $25.54\errUD{13.30}{10.63}$ & $1.52\errUD{1.24}{0.93}$ & $3.11\errUD{1.64}{1.21}$ & $20.90\errUD{10.42}{8.49}$ & $2.07$ & $0.88$ & $0.39$ & $0.80$ \\
    {\it (B)} & $33.53\errUD{28.25}{18.30}$ & $2.74\errUD{2.30}{1.58}$ & $4.16\errUD{4.12}{2.42}$ & $26.63\errUD{21.83}{14.31}$ & $2.61$ & $1.18$ & $0.44$ & $1.00$ \\
    {\it Nov.~2002} & $15.14\errUD{10.74}{5.60}$ & $0.37\errUD{0.69}{0.17}$ & $0.94\errUD{0.37}{0.56}$ & $13.84\errUD{5.04}{9.08}$ & $0.96$ & $0.14$ & $0.19$ & $0.63$ \\
    \cline{1-9}          
  \end{tabular}
}
\end{center}
{\scriptsize Errors are quoted at the 90\% confidence level for 1 parameter of interest ($\Delta\apix{\chi}{2}=2.71$); narrow line energy fixed to $\pedSM{E}{rf}=6.4\;$keV. For the accretion disk reflection component, the outer radius was fixed to the maximum value, $\pedSM{R}{out}=400\,\pedSM{r}{g}$. $^{\mydag}\;$For the single state, the power law and the accretion disk reflection components have the same photon index. $^{\myddag}\;$Parameters for the states tied in the fit. $^a\;$In units of $10^{-4}\,$\normPL\,@$1\,$keV. $^b\;$In units of $10^{22}\;$\nh. $^c\;$Ionization parameter $\xi \equiv 4\pi \pedSM{F}{ill}/n$ (erg~cm~s$^{-1}$), where \pedSM{F}{ill} is the ionizing flux and $n$ is the hydrogen nucleus density (part~cm$^{-3}$) of the illuminated slab. $^b\;$In units of \pedSM{r}{g}. $^d\;$Complement to one of the covered-to-uncovered flux ratio. $^e\;$Ratio between the ionized reflection flux and the total flux in the $(0.1-1000)\,$keV energy range. }  \\
\end{minipage}
\end{table*}

\vspace{0.5cm}
Regarding the two-phase warm absorber-based scenario, to improve the fit we added to the power law (covered by an ionized, partial covering absorber and an ionized, fully covering absorber) a soft power law component absorbed only by neutral gas ($\Delta\apix{\chi}{2}/\Delta\mbox{d.o.f.}=14.9/1$). 
If we let the soft power law photon index to be a free parameter, the fit improves significantly but still remains statistically poor ($\chidof=350.0/279$). 
Looking at the residuals, it seems that the ABSORI model is inadequate to describe the complex shape of the absorbing structure seen at $0.7-1\;$keV. 
Moved by these considerations, we try to reproduce the soft complexities with the cited XSTAR tables.

In order to apply these tables, for this part of the analysis we passed to XSPEC version~11.
We fitted to the data our model (a double ionized absorber, one partially covering and the other totally covering the source, plus a soft scattered component) using a pre-calculated XSTAR table.
The publicly available table ``grid 18b'' assumes a $10^{44}\;$\lum, $\Gamma=2$ power law ionizing a constant density shell of gas whose abundances can be varied. 
The turbulent velocity of gas is assumed to be $100\;$km~s$^{-1}$. 
We obtain a good fit to the data with an oxygen overabundance of $\sim 4\times$solar, requested to reproduce the deep absorption structure observed between $0.7$ and $1\;$keV.
We note here that an ionizing continuum steeper than $\Gamma=2$ results in deeper resonant absorption in this energy range due to the higher number of soft ionizing photons \citep[see \eg][]{nicastro99}.
If a grid generated assuming a continuum with $\Gamma=2.5$ is applied to the model, no oxygen overabundance is needed to reproduce the depth of the absorption structure; however, in this case the enhanced opacities at low energies ($E<0.7\;$keV) cannot reproduce the flat spectra observed in this energy range. 
We decided to proceed with the analysis adopting the ``grid 18b'' table, with these considerations in mind.

A visual inspection of data/model ratio shows an excess of emission in the soft band. 
The addition of a gaussian emission line improves the fit. 
The line energy, $E\sim 0.59\;$keV, is fully compatible with the photoionized \ion{O}{vii} triplet energy.
Soft X-ray emission lines, mainly \ion{O}{vii} or \ion{Ne}{ix}, have been found in the spectrum of other Seyfert galaxies (\eg, \object{Mrk~6}, \citealt{immler03}; \object{Mrk~1239}, \citealt{grupe04abs}; \object{Mrk~304}, \citealt{piconcelli04}; \citealt{guainazzi07} and references therein).
Broad components of the \ion{O}{vii} triplets associated with the BLR have been observed in the X-ray spectra of the bright NLS1 \object{Mrk~110} \citep{boller07} and of the Seyfert galaxy \object{Mrk~279} \citep{costantini07}.
Unfortunately, the flux of \sorg\ is too low and prevents us from drawing information from the RGS spectra.

The best fit model results in a moderately ionized gas ($\nhsym\sim 8.4\times10^{22}\;$\nh, $\xi\sim 19.9\;$erg~cm~s$^{-1}$) covering $\sim 40$\% of the primary emission, a second absorber with lower column density and higher ionization state ($\nhsym\sim 4.5\times10^{22}\;$\nh, $\xi\sim 94\;$erg~cm~s$^{-1}$) fully covering the source, a scattered component and two gaussian emission lines (Fe and \ion{O}{vii}).
The absorbed power law is rather steep ($\Gamma \sim 2.5$), and the scattered one is even steeper, as expected if scattering occurs in a photoionized gas. 
Finally, the whole emission is covered by a neutral absorber with a rather low column density ($\nhsym\sim 2\times10^{21}\;$\nh).
This complex model results in an equally good description of the (A) spectra ($\chi^2$/dof$=297.5/279$, see second part of Table~\ref{tab:broad}).
For comparison, in Fig.~\ref{fig:ufde23} we have plotted the unfolded best fits to the pn data obtained with a partially absorbed accretion disk reflection ({\it left panel}), or a combination of ionized absorbing layers and the addition of scattered emission ({\it right panel}).

A partially ionized absorbing material with large velocity shear (the SWIND\footnote{http://heasarc.gsfc.nasa.gov/docs/xanadu/xspec/models/swind1.html} model in XSPEC, originally proposed by \citealt{swind04} and updated by \citealt{swind06}) as origin of the observed soft excess cannot be tested here, being the maximum column density assumed in the analytical model ($\nhsym\sim 8\times10^{22}\;$\nh) lower than the value we need to reproduce the high energy spectral shape.

%
\begin{table*}
\begin{minipage}[!h]{2\columnwidth}  
\caption{\footnotesize Simultaneous analysis of the three available \xmm\ datasets for \sorg\ in the $0.4-10\;$keV energy range: the adopted model was the complex absorption scenario: an ionized, partially covering absorber plus a scattered component emerging from an ionized layer that fully covers the source; $\chidof=417.5/376$).
Luminosity are absorption-corrected, while fluxes are corrected only for the Galactic absorption.}
\label{tab:allbfAbs}
\begin{center}   
\renewcommand{\footnoterule}{}  
{\tiny
\begin{tabular}{r@{\extracolsep{0.18cm}} c@{\extracolsep{0.13cm}} c@{\extracolsep{0.18cm}} c@{\extracolsep{0.13cm}} c@{\extracolsep{0.18cm}} c@{\extracolsep{0.13cm}} c@{\extracolsep{0.13cm}} c@{\extracolsep{0.18cm}} c@{\extracolsep{0.13cm}} c@{\extracolsep{0.13cm}} c}
 \vspace{-0.2cm}  \\            
  \hline\hline  
   & \multicolumn{2}{c}{Power law$^{\mydag}$} & \multicolumn{3}{c}{PC warm abs.} & \multicolumn{2}{c}{Warm abs.} & \multicolumn{2}{c}{Gaussian} & Neutral abs. \\
  \cline{2-3} \cline{4-6} \cline{7-8} \cline{9-10} \cline{11-11}
   Obs.
   & $\Gamma$
   & Norm$^a$
   & \nhsym$^b$
   & $\xi$$^c$ 
   & Cov.~fraction$^d$
   & \nhsym$^b$
   & $\xi$$^c$ 
   & \pedSM{E}{rf}$^e$
   & $\sigma$$^f$
   & \nhsym$^b$ \\
   & \pedSM{\Gamma}{scatt}
   & Scatt.~fraction$^g$
   & 
   & 
   & 
   & 
   & 
   & 
   & 
   & \\
\hline
   {\it (A)} & $2.54\errUD{0.63}{0.49}$ & $32.56\errUD{12.00}{5.17}$ & $8.24\errUD{5.53}{1.12}$ & $18.16\errUD{7.23}{2.18}$ & $0.62\pm 0.03$ & $3.95\errUD{0.42}{0.23}$ & $91.32\errUD{2.81}{2.03}$ & $0.56\pm 0.02$ & $40\pm 30$ & $0.22\errUD{0.05}{0.01}$ \\
             & $2.90\errUD{0.65}{0.47}$ & $0.005\errUD{0.004}{0.002}$ &  &  &  &  &  &  &  &  \\
   {\it (B)} & $2.54^{\myddag}$ & $37.33\errUD{6.56}{7.77}$ & $8.24^{\myddag}$ & $18.16^{\myddag}$ & $0.51\errUD{0.04}{0.03}$ & $3.95^{\myddag}$ & $91.32^{\myddag}$ & $0.56^{\myddag}$ & $40^{\myddag}$ & $0.22^{\myddag}$ \\
             & $2.90^{\myddag}$ & $0.004\pm 0.001$ &  &  &  &  &  &  &  &  \\
   {\it Nov.~2002} & $2.54^{\myddag}$ & $30.60\errUD{13.43}{4.34}$ & $8.24^{\myddag}$ & $18.16^{\myddag}$ & $0.91\errUD{0.03}{0.04}$ & $3.95^{\myddag}$ & $60.69\errUD{10.10}{23.56}$  & $0.56^{\myddag}$  & $40^{\myddag}$ & $0.22^{\myddag}$ \\              
             & $2.90^{\myddag}$ & $0.007\errUD{0.002}{0.003}$ &  &  &  &  &  &  &  &  \\
  \hline
  \multicolumn{11}{l}{} \\                 
    & \multicolumn{4}{c}{Observed flux $[10^{-13}\,\flux]$} & \multicolumn{4}{c}{Unabsorbed luminosity $[10^{43}\,\lum]$} \\     
    \cline{2-5}\cline{6-9}          
   Obs.
    & $0.4-10\;$keV
    & $0.4-1\;$keV
    & $1-3\;$keV
    & $3-10\;$keV 
    & $0.4-10\;$keV
    & $0.4-1\;$keV
    & $1-3\;$keV
    & $3-10\;$keV \\
    \cline{1-9}          
    {\it (A)} & $25.30\errUD{5.55}{0.59}$ & $1.53\pm 0.01$ & $3.18\errUD{0.18}{0.02}$ & $20.59\errUD{5.38}{0.57}$ & $5.70$ & $2.72$ & $1.87$ & $1.11$ \\
    {\it (B)} & $33.15\errUD{2.48}{1.48}$ & $2.60\pm 0.01$ & $4.48\errUD{0.09}{0.03}$ & $26.06\errUD{2.41}{1.43}$ & $7.24$ & $3.55$ & $2.31$ & $1.38$ \\
    {\it Nov.~2002} & $14.93\errUD{1.54}{2.12}$ & $0.35\errUD{0.01}{0.02}$ & $0.94\errUD{0.01}{0.06}$ & $13.64\errUD{1.54}{2.04}$ & $4.22$ & $2.02$ & $1.38$ & $0.82$ \\
    \cline{1-9}          
\end{tabular}
}
\end{center}
{\scriptsize Errors are quoted at the 90\% confidence level for 1 parameter of interest ($\Delta\apix{\chi}{2}=2.71$); Fe narrow line energy fixed to $\pedSM{E}{rf}=6.4\;$keV, width fixed to $0\;$eV. $^{\mydag}\;$First line, intrinsic power law; second line, scattered power law. $^{\myddag}\;$Parameters for the states tied in the fit. $^a\;$In units of $10^{-4}\,$\normPL\,@$1\,$keV. $^b\;$In units of $10^{22}\;$\nh. $^c\;$Ionization parameter $\xi \equiv 4\pi \pedSM{F}{ill}/n$ (erg~cm~s$^{-1}$), where \pedSM{F}{ill} is the ionizing flux and $n$ is the hydrogen nucleus density (part~cm$^{-3}$) of the illuminated slab. $^d\;$Complement to one of the covered-to-uncovered flux ratio. $^e\;$Line centroid energy, in keV. $^f\;$Line width, in eV. $^g\;$Ratio between the scattered flux and the partially absorbed component. }  \\
\end{minipage}
\end{table*}
%


\section{The origin of X-ray variability: comparison among the different \xmm\ datasets}\label{sect:var}

The physical scenarios explored before to describe the (A) data yielded two complex models, composed by several continuum components and different absorbers.
Both models are statistically acceptable.
We then explored the possibility to explain the variability observed between \dataO\ and \dataN\ (both in shape and intensity) by changing only the best fit parameters (but assuming the same components in the model): we applied the models described in \S\ref{sect:wholeA} to the \dataO\ and (B) datasets.
The two models provide a good description for the two spectra with reasonable values for the model parameters.
In the absorption-based model, the soft gaussian emission line at $\sim 0.57\;$keV is not requested by the \dataO\ dataset: its EW upper limit is $273\;$eV at 90\% confidence level. 

To conclude the analysis, we tested the possibility of fitting these models to the three states simultaneously, changing as less parameters as possible.

The large number of variables involved in the reflection-based model (some of them degenerate in the fit) can make difficult the identification of the parameters responsible of the observed variations.
To choose the parameters to be free, we started by fitting the spectra simultaneously with all but the normalization tied toghether.
We then thawed one parameter at a time, searching for an improvement in the fit.
The same step was repeated with an increasing number of free parameters, in various combinations, until the improvement stops to be significant (\ie, $F$-test probability lower than $95$\%).
In the final fit, the parameters identified in this way were kept free, while we tied the others between the states.
Variations in the inner ionized absorber, partially covering the central source, coupled with an increase of the reflection fraction, produce the differences observed between (A) and (B) spectra.
To explain the change between \dataO\ and \dataN, both absorbing layers must vary; a lower accretion disk ionization state in the oldest observation is also required.

The absorption-based model can reproduce the three spectra by varying three parameters: the intrinsic power law normalization, the covering fraction of the inner absorber and the ionization parameter of the outer absorber.
The power law normalization varies roughly proportionally to the observed hard ($3-10\;$keV) flux.
Variations observed between (A) and (B) datasets can be explained by means of a different covering fraction of the inner ionized absorber. 
The differences between \dataO\  and \dataN\ spectra can be attributed to a different covering fraction of the inner absorber plus a variation of the ionization state of the outer absorber.

\begin{figure}[!b]
  \centering
  \includegraphics[height=4.4cm,width=4.35cm,angle=270]{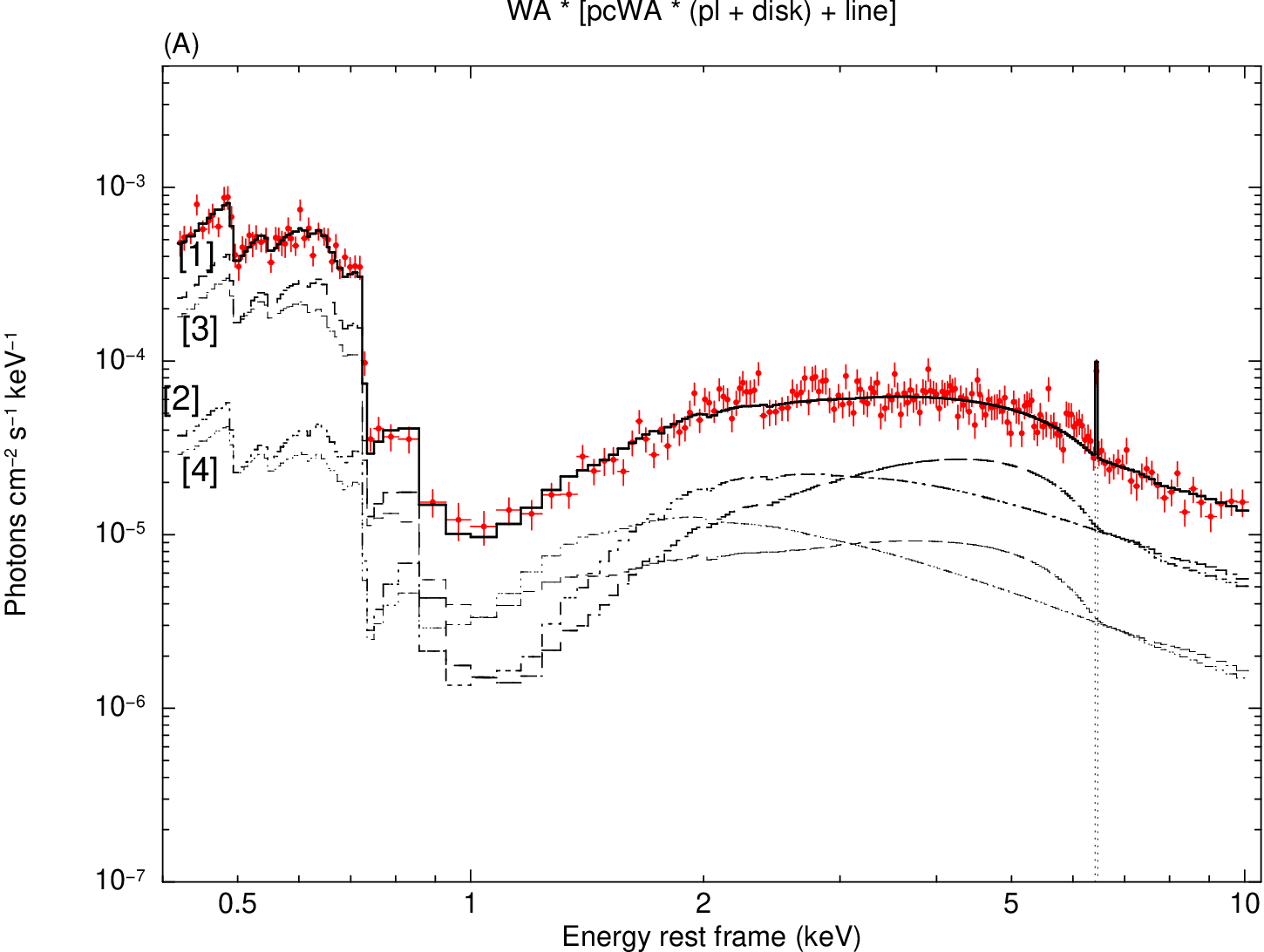}
  \includegraphics[height=4.4cm,width=4.35cm,angle=270]{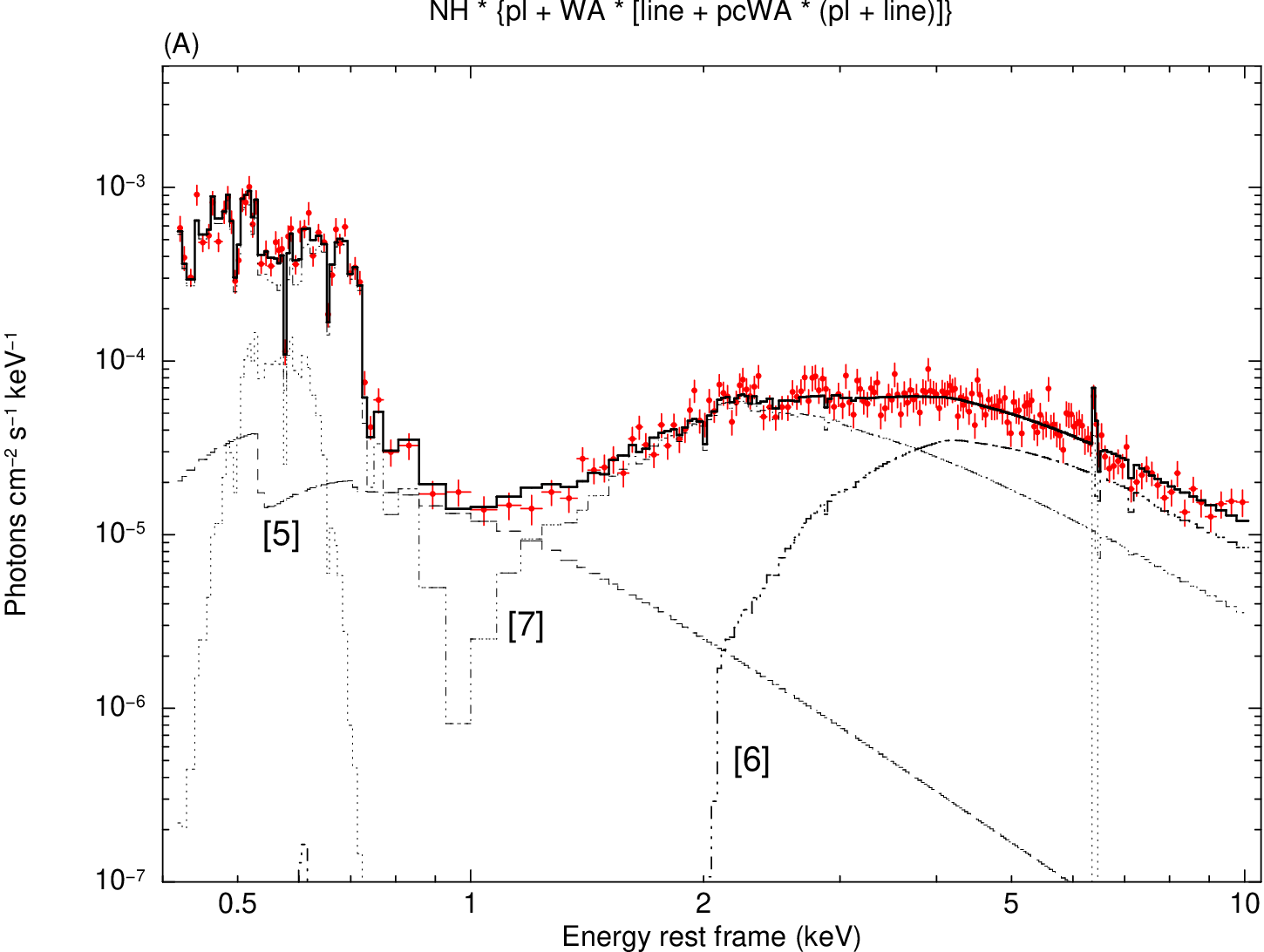}
  \includegraphics[height=4.4cm,width=4.35cm,angle=270]{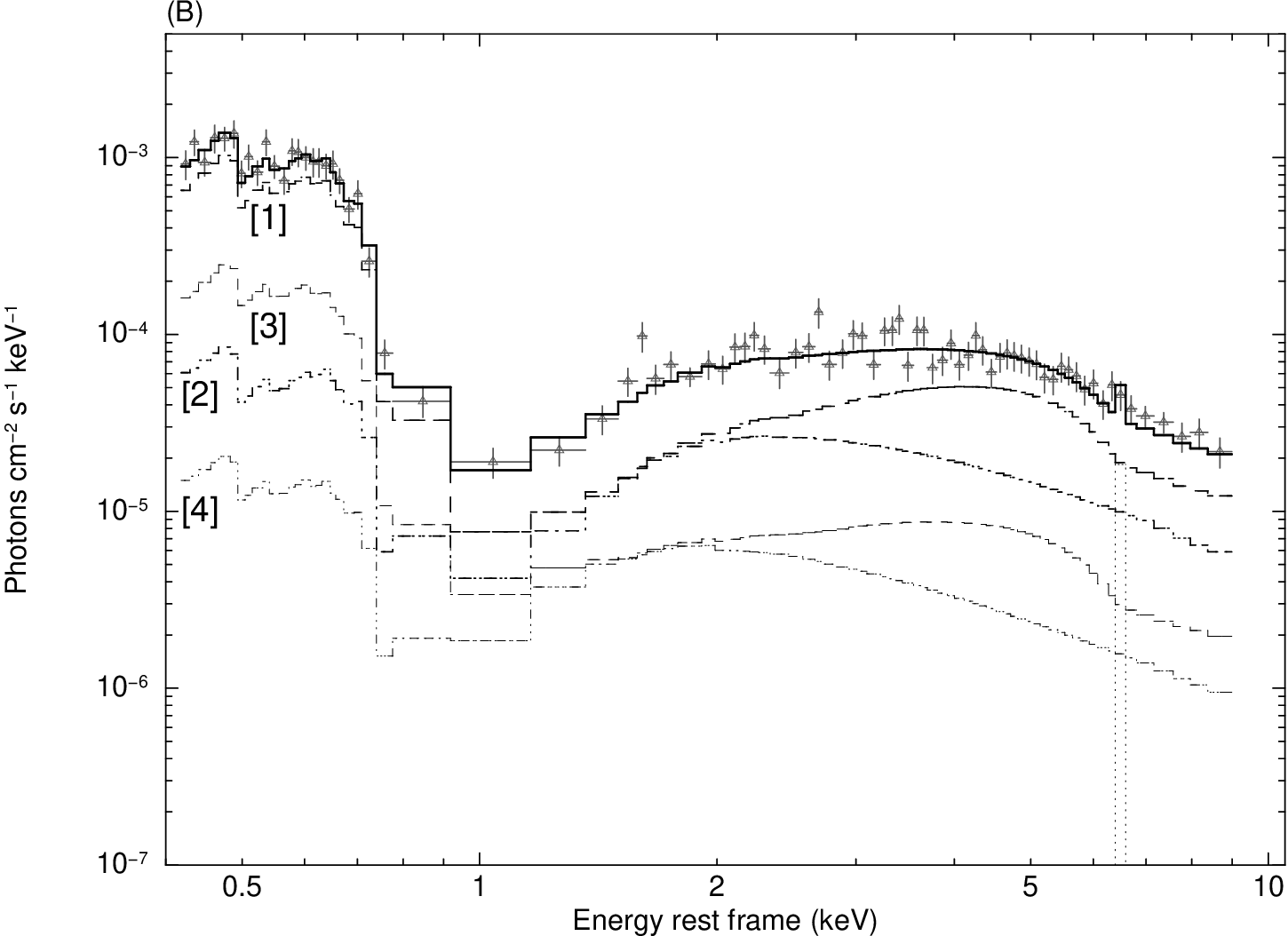}
  \includegraphics[height=4.4cm,width=4.35cm,angle=270]{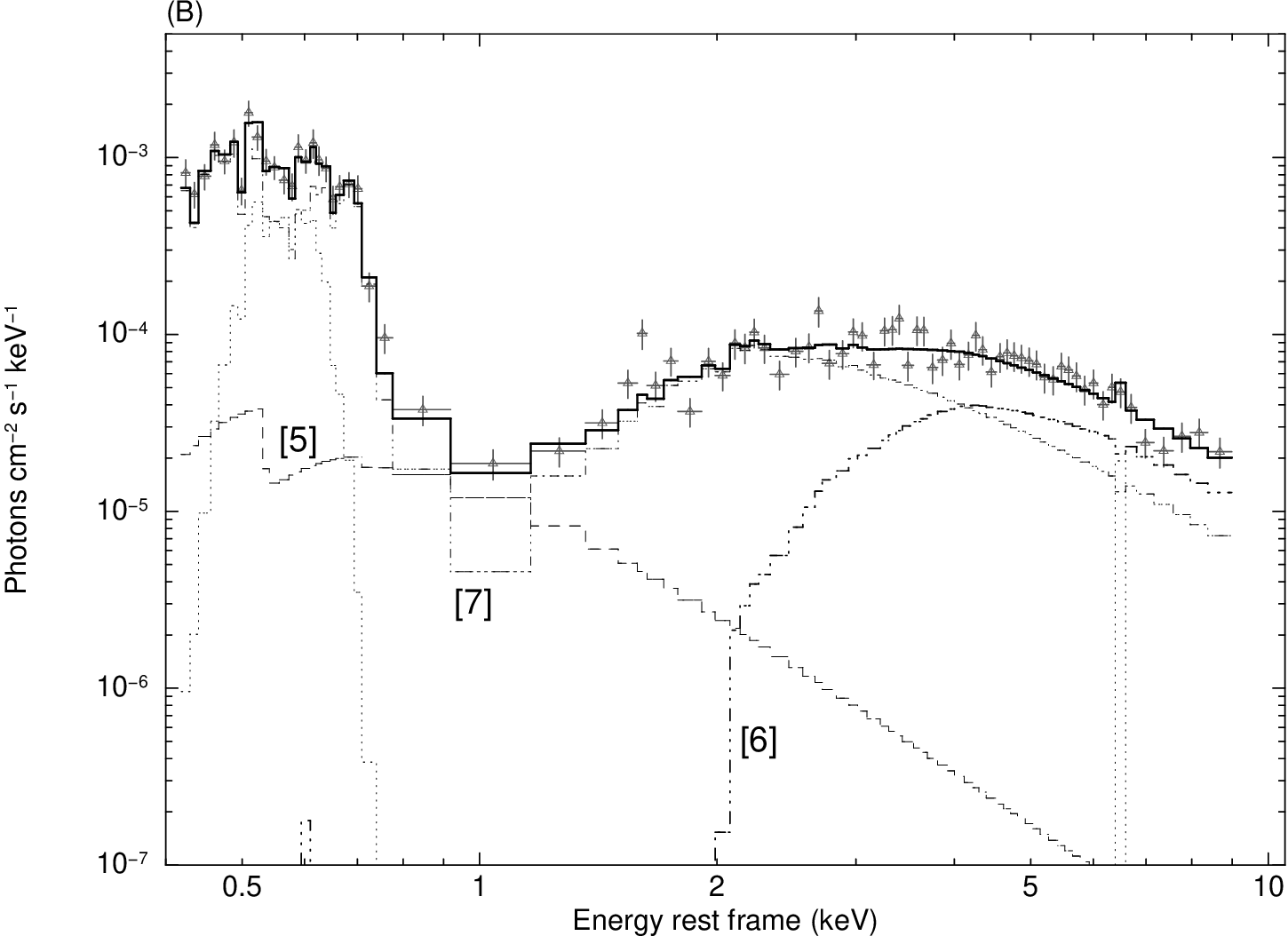}
  \includegraphics[height=4.4cm,width=4.35cm,angle=270]{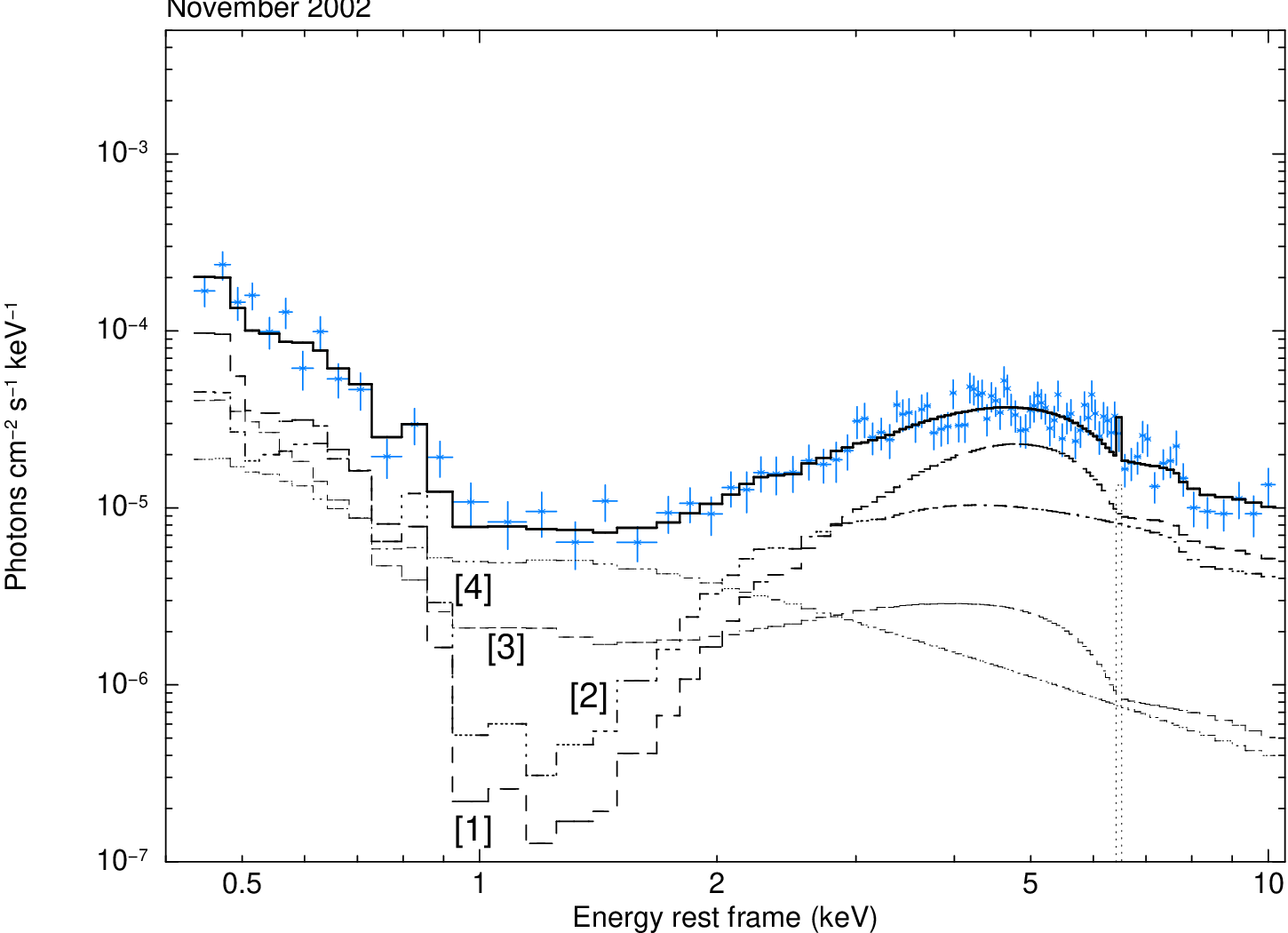}
  \includegraphics[height=4.4cm,width=4.35cm,angle=270]{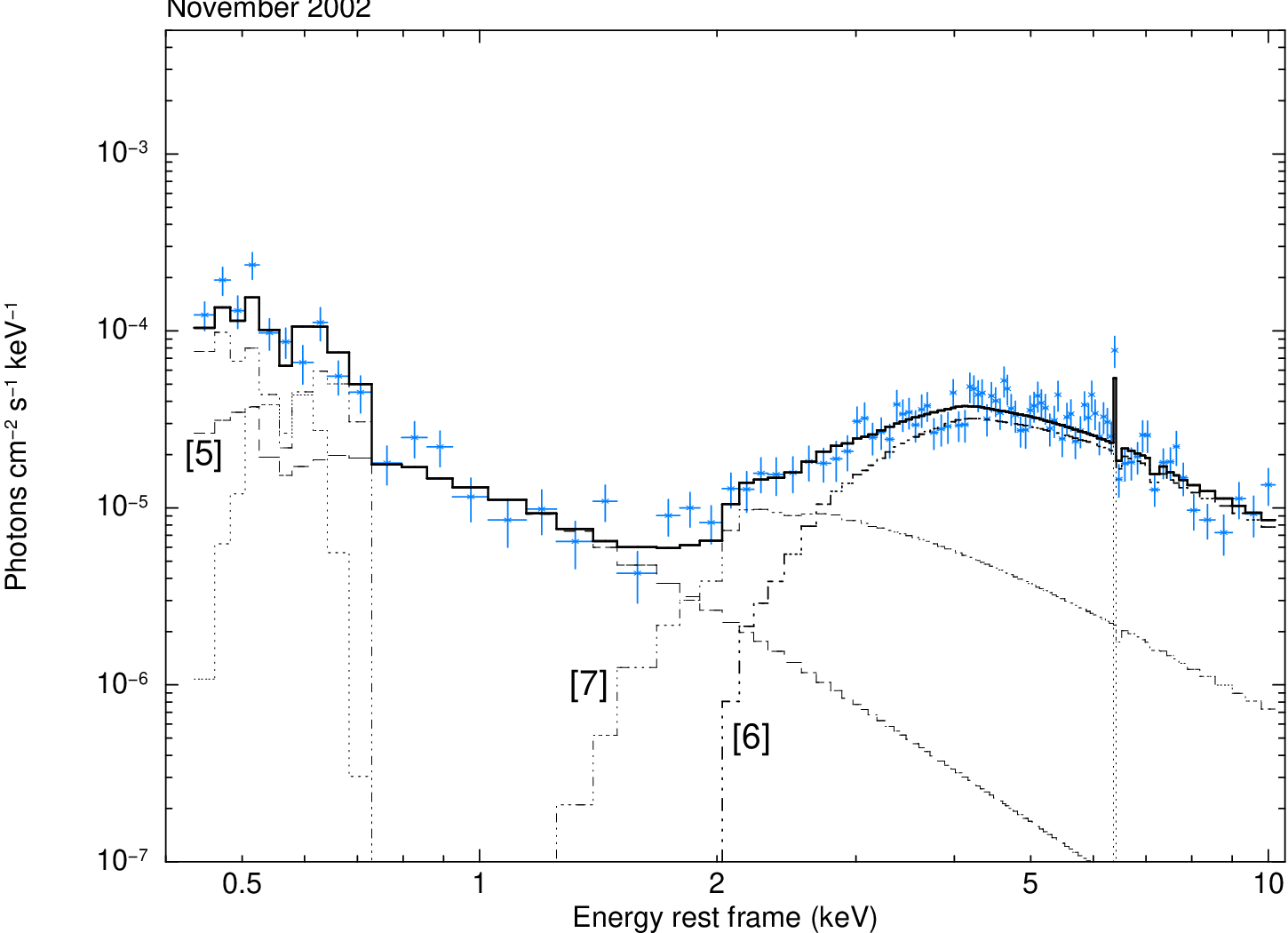}
  \caption{Unfolded EPIC pn spectra of \sorg\ with the two adopted models fitted simultaneously to the three observations (see \S\ref{sect:var}): the \dataN\ (A) spectra (filled circles and red lines, {\it upper panels}) and (B) spectra (open triangles and gray lines, {\it middle panels}), and the \dataO\ spectra (stars and sky-blue lines, {\it lower panels}).
           The solid lines correspond to the total best fit models.
	   Over-plotted are the different components of each model: {\it left panels}, reflection from the disk (dashed line), power law (dash-dotted line), and narrow emission line (dotted line), with the thickest lines identifying the fraction covered by the partial covering warm absorbed ([1] and [2] respectively), while the thinnest lines identify the fraction uncovered, ([3] and [4] respectively); {\it right panels}, scattered component (dashed line, [5]), absorbed power law (dash-dotted line) and emission lines (dotted line), with the thickest lines identifying the fraction covered by the partial covering warm absorber, [6], and the thinnest the fraction uncovered, [7].}
  \label{fig:alluf}%
\end{figure}

The adopted models are able to describe the three states in a consistent way ($\chidof=379.6/364$ and $417.5/376$; see Fig.~\ref{fig:alluf}): the best fit value for the parameters are reported in Table~\ref{tab:allbfDisk} and Table~\ref{tab:allbfAbs}.

%
%


\section{Summary and Discussion}

We have presented a detailed spectral analysis of all the EPIC \xmm\ data available of \sorg.
The source has been first observed in \dataO, and then re-observed in \dataN; the latter data are presented here for the first time.

\subsection{X-ray emission}

Several features are clearly evident in the spectra of \sorg\ (see Fig.~\ref{fig:obs23pl}): a strong absorption between $\sim 0.7$ and $2\;$keV, a spectral flattening at lower energies, a complex spectral shape in the iron band, and a drop in the emission at higher energies.
Moreover, strong variations in shape and intensity are evident both on short and long timescales.
The flux at energies $<1\;$keV increases of a factor $\sim 10$ in three years, then decreases of a factor $\sim 2$ in six days.
Above $3\;$keV the variability is significantly less important, with a change in flux of a factor $\sim 2$ and $\sim 1.3$ in the two time intervals, respectively.

The three broad band ($0.4-10\;$ keV) EPIC spectra were analysed in the framework of two different models: an ``absorption-based'' scenario and a ``relativistic reflection-based'' one.
In both cases, a warm absorber is detected, producing the strong absorption features observed in the $0.7-2\;$keV energy range. 
The spectral curvature observed at high energy can be reproduced either by a partial covering absorber or by a blurred relativistic reflection component. 
In both cases, also a narrow gaussian emission line, consistent with K$\alpha$ fluorescence from almost neutral iron, is detected. 

To account for the observed spectra assuming an accretion disk reflection contribution, the continuum emission must be intercepted by two layers of matter.
The innermost, ionized absorber covers a high fraction (between $\sim 0.7$ and $\sim 0.9$) of the emission.
The second layer, fully covering the source, is almost neutral ($\xi\sim 3.3\;$erg~cm~s$^{-1}$).
As noted, this model suggest a reflection produced in regions penetrating deeply towards the central blak hole ($\pedSM{R}{in} < 2.50\;$\pedSM{r}{g}).
The disk emissivity power law index is close to $4$, in agreement with theoretical expectations in case of a small height of the primary source \citep[\eg,][]{martocchia02}.
The reflection fraction, defined as the ratio between the flux in the ionized reflection component and the total flux, is of the order of $50-60$\%, relatively large with respect to the $20-30$\% expected in the case of an isotropic illumination.
The large amount of reflection from the disk observed in all states requires some mechanism to enhance the reflection component with respect to the intrinsic continuum, resulting in a strongly suppressed intrinsic emission.
Recent theoretical works, based on the ``light bending'' effect \citep{miniutti03,miniutti04}, or an inhomogeneous structure of the accretion disk due to violent clumping instabilities \citep{merloni06}, predict this kind of effect.
A similar spectral behaviour has been recently observed in the \xmm\ spectra of \object{Mrk~841} \citep{petrucci07}.

The main origin of the observed variability results to be a change in the absorbers properties (column densities and covering fractions), coupled with a change in the reflected component and the disk ionization state (see Table~\ref{tab:allbfDisk}).
The coverage of the inner absorber decreases passing from \dataO\ to (B) and (A) spectra.
A higher disk reflection fraction and a higher coverage of the inner absorber explain the spectral changes observed passing from (A) to (B) states.
The spectral index of the intrinsic continuum ($\Gamma\sim 2$) is consistent within the errors with the value found for radio-quiet AGN \citep[$\Gamma \sim 1.9$;][]{piconcelli05,nandra07} and the mean value found for other NLS1s in the $2-10\;$keV energy range \citep[$\Gamma \sim 2.19$;][]{leighly99}.
The differences between \dataO\ and \dataN\ spectra imply stronger variations.
The primary continuum changes both in shape (flatter in the oldest observation, $\Gamma\sim 1.8$) and in intensity, and the accretion disk ionization state increases.
The column densities of the two absorbers change in an opposite way: in \dataO\ the column density of the inner absorber is higher ($\nhsym\sim 2\times 10^{23}\;$\nh), while the outer absorber shows a column density lower than in the \dataN\ spectra ($\nhsym\sim 5\times 10^{21}\;$\nh).

Finally, a narrow \feka\ emssion line is included in the model, probably produced by the obscuring torus postulated in AGN Unified Models (but other possible sites are the outer part of the accretion disk, or the optical BLR).
Recently, a narrow emission line associated with cold iron was found to be almost ubiquitous in a sample of local radio-quiet Seyfert galaxies observed with \xmm\ \citep{nandra07}.
However, $\sim 70$\% of these objects present further complexities in the iron band.
Their spectral analysis shows that a model based on broad emission from an accretion disk is the best way to take into account this complexity.
\citet{elena05} noted that a narrow \feka\ emission line is required in $\sim 53$\% of the PG~AGN sample studied by \citet{piconcelli05}.
The detection of broad relativistic lines is statistically significant in less than $10$\% of their sources (namely \object{PG~0007+106}, \object{PG~0050+124} and \object{PG~1116+215}).
Interestingly, their hard X-ray luminosity falls in the lower end of the range covered by the sample ($\pedix{L}{2-10\kev}<10^{44}\;$\lum): in the disk reflection framework, the unabsorbed $2-10\;$keV luminosity observed from \sorg\ changes from $9.7\times10^{42}\;$\lum\ to $1.2\times10^{43}\;$\lum\ and $7.2\times10^{42}\;$\lum\ for the (A), (B) and \dataO\ spectra, respectively.
From their \xmm\ data analysis of a large sample of AGN, \citet{guainazzi06} found that relativistically broadened \feka\ lines are significantly more common in low luminosity ($\pedix{L}{2-10\kev}\leq 10^{43}\;$\lum) AGN.

In the absorption-based scenario several components are needed to properly reproduce the data: the primary power-law emission is partially covered by the warm absorber above mentioned, and totally covered by a second ionized absorber. 
A soft scattered component is also present. 
In addition to the \feka\ line, an \ion{O}{vii} emission line (not required in the reflection-based model) is also detected at $E\sim 0.57\;$keV.
Finally, the whole emission is absorbed by a rather low ($\sim 2\times 10^{21}\;$\nh) column density of neutral gas.

The observed variability can be accounted for by variations of the two warm absorber components. 
Between \dataO\ and \dataN\ observations, the covering fraction of the inner absorber decreased significantly (passing from $\sim 0.9$ to $\sim 0.55$), leading to a different shape of the complex curvature seen at high energies. 
In the meanwhile the ionization state of the outer absorber increased (from $\sim 60$ to $\sim 90\;$erg~cm~s$^{-1}$) in a manner roughly proportional to the increase in unabsorbed luminosity. 
This may suggest that the density and location of the warm absorber has not changed in this lapse of time.
Instead, the differences observed between (A) and (B) spectra can be explained by a variation of the covering fraction of the inner ionized absorber.

The inner, partially covering warm absorber presents a lower ($\xi\sim 20\;$erg~cm~s$^{-1}$) ionization state than the fully covering warm absorber. 
In accretion disk winds theoretical models \citep[\eg,][]{proga04}, two phase of ionized gas are predicted to coexhist in the inner regions of accretion disks, where the wind is ``launched''. 
In particular, the wind itself is shielded from the central ionizing source by a higher ionization component. 
In this framework, it is possible that we are looking \sorg\ at the base of the wind, with the partial covering absorber corresponding to the ``wind blobs'' and the totally covering absorber corresponding to the highly ionized shielding phase. 
The latter warm absorber component may be responsible for the scattering of the primary emission into our line of sight.
The very small scattering fraction (few parts per thousand) could thus be explained by the low column density ($\sim 4\times 10^{22}\;$\nh) of the scatterer, coupled with a possible anisotropic distribution around the source.
In this scenario, the emission line at $\sim 6.4\;$keV can be produced by transmission through the high column density, partially covering absorber, while the line detected at $\sim 0.6\;$keV could be associated to photoionized emission occurring in the low column density, totally covering absorber. 
The presence of unresolved photoionized emission lines would then naturally explain the steepening of the scattered power law with respect to the intrinsic one.
The primary power law emission is somewhat steep ($\Gamma\sim 2.5$), but consistent within the errors with the values found by \citet{leighly99}.
Finally, the small ($\nhsym\sim 2\times 10^{21}\;$\nh) neutral column density absorbing the whole emission and affecting only the data below $0.5\;$keV could be ascribed either to dust lanes in the host galaxy or to the presence of dust in the wind and/or scattering region.
Overall, the complex spectra of \sorg\ seem thus fully compatible with the presence of a strong and complex ionized outflow along the line of sight.

\subsection{The broad-band SED of \sorg}

Considering the strong variability observed in the X-ray band, it is interesting to look at the multiwavelength emission properties of \sorg. 

In the UV band, \citet{gallagher01} noted a variation of $30$\% between the \iue\ data of November 1982 and May 1984, and earlier \iue\ observation from June 1982 and September 1982, that are found consistend with later \hst\ spectra.
The fluxes observed by the OM simultaneously to the new EPIC spectra presented here are in good agreement with the \hst\ data.

At long wavelengths, \sorg\ is the only radio-quiet AGN observed by \citet{neugebauer99} where a variation in the near-infrared (NIR) emission from 1980 to 1998 is reported: in particular, the flux in $4$ bands from $1.27\:\mu$m to $3.7\:\mu$m rises of $0.5\;$mag between 1993 and 1996, going back to the previous value between 1996 and 1998.
The continuum emission observed in Seyfert galaxies at $\lambda\gtrsim2\:\mu$m is very likely dominated by thermal radiation from the inner edge of the dust torus \citep[\eg,][]{kobayashi93}, heated by the radiation originated in the central active nucleus \citep{haas03}.
How much the torus extends towards the black hole depends on the irradiating luminosity and on the dust composition, that determine the distance at which the temperature is so high that the dust begins to sublimate.
So, the dust evaporation radius can be estimated from the bolometric luminosity, $\pedix{R}{evap} \simeq 0.06 \sqrt{\pedix{L}{bol}/(10^{45}\lum)}\;$pc \citep{netzer93,hill96}.
Assuming that the inner radius of the torus roughly corresponds to \pedix{R}{evap}, and adopting the bolometric luminosity estimated from the \OIII\ line width by \citet{zhang06}, $\pedix{L}{bol}\sim 1.4 \times 10^{45}\;$\lum, we can estimate the dimension of the region producing the observed NIR emission, $\pedix{R}{torus}\sim 0.07\;$pc.
The minimum timescale for an observable variability implied by \pedix{R}{torus} is of the order of a year.
Therefore, it is consistent with the possibility that the observed NIR variability is the reaction of the torus to the long timescale X-ray variable emission.

In \S\ref{sect:sorg} we have mentioned the high optical continuum polarization showed by \sorg, $P=2.5$\% \citep{berriman90}.
The optical polarization spectrum presented by \citet{smith97} exhibits a rise in the linear polarization $P$ toward shorter wavelength (roughly proportional to $\lambda^{-1}$), that continues into the UV band down to $2000\;$\AA.
A complex behaviour of the polarization position angle across the \ha\ feature is also observed.
To explain the continuum and line polarizations, the authors suggest the presence of different scattering regions, with at least part of the scattering material intermixed with the line-emitting gas.
Warm electrons responsible for the observed polarization in the emission-line wings (\ie, the interaction with the warm absorber required by the X-ray observations) would be a possible alternative.
Considering the two-component scattering geometries recently proposed to account for the polarization properties of all Seyfert galaxies simultaneously \citep{smith02,smith04}, we can think that we are observing \sorg\ at a ``privileged'' angle.
Although the available informations do not allow \citet{smith04} to confidently assign a polarization class (``polar-'' or ``equatorially-'' scattered Seyfert) to this object, it is possible that our line of sight passes very close to the absorbing torus.
This seems to be a coherent scenario that take into account observations in different bands; further multiwavelength observations are required to test this hypothesis.

%
\begin{figure}[t]
  \centering
  \resizebox{\hsize}{!}{\includegraphics[angle=90]{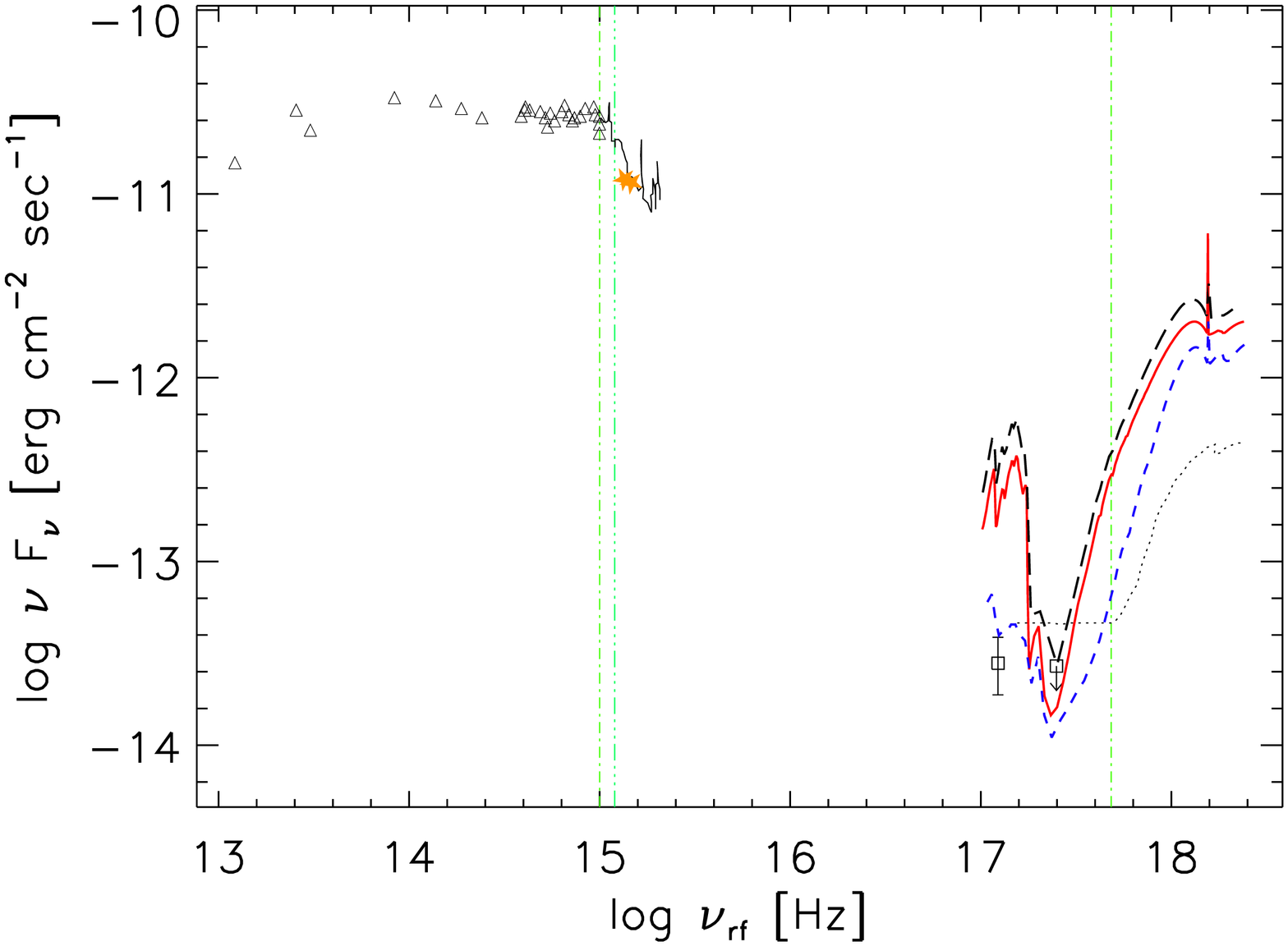}}
  \caption{Spectral energy distribution for \sorg.
           At low frequencies, open triangles are infrared and optical continuum data from \citet{neugebauer87}, and the thin solid curve is the UV spectrum from \hst\ \citep{gallagher01}.
	   Yellow filled stars indicate the \xmm\ OM fluxes observed in \dataN.
	   In the X-ray energy range, \rosat\ data point and upper limit are marked with open squares, while the black dotted line indicates the \asca\ best fitting model \citep{gallagher01}.
	   The three observed \xmm\ EPIC fluxes are shown with the black long dashed line, \dataN\ spectrum (B); red solid line, \dataN\ spectrum (A); blue dashed line, \dataO\ spectrum.
	   The vertical thin dot-dashed light-green lines correspond to $\pedix{\lambda}{rf}=3000\;$\AA\ and $\pedix{E}{rf}=2\;$keV, while the green dot-dot-dot-dashed line corresponds to $\pedix{\lambda}{rf}=2500\;$\AA.}
  \label{fig:multinused}%
\end{figure}

In Fig.~\ref{fig:multinused} we present the broad-band SED of \sorg.
A stronger high-energy variability than at low frequencies is clearly evident.
The light-green dot-dashed lines corresponds to $\lambda=3000\;$\AA\ and $E=2\;$keV in the source rest frame, \ie\ where the low and high energy fluxes used by \citet{brandt00} to evaluate the \pedix{\alpha}{ox} are calculated.
The change observed in the X-ray flux in front of a roughly constant UV flux implies a variability in the X-ray--to--optical index even higher than previously reported: \citet{gallagher01} estimated for \rosat\ and \asca\ observations $\pedix{\alpha}{ox}<-2.17$ and $\pedix{\alpha}{ox}=-2.03$.
The {\it observed} indices measured from the new \xmm\ spectra are considerably higher: we can compute the optical fluxes rescaling the UV \hst\ spectrum to the OM data (that, we stress, are simultaneous to the X-ray fluxes), obtaining $\pedix{\alpha}{ox} = -1.66$ and $-1.65$ for the (A) and (B) spectra.
This result implies a change in the classification of \sorg\ as X-ray weak source. 

%
\begin{figure}[t]
  \centering
  \resizebox{\hsize}{!}{\includegraphics[width=8cm]{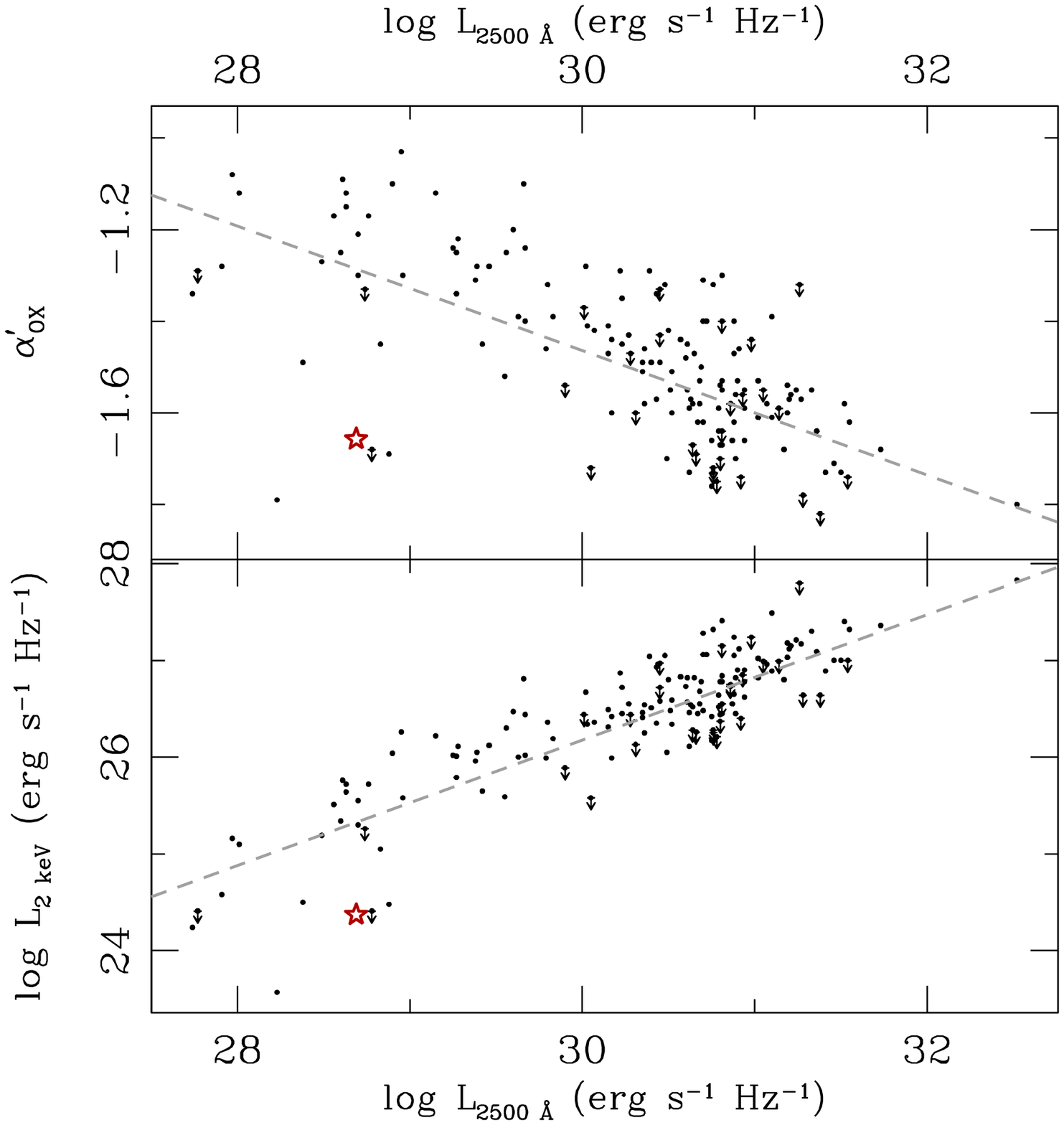}}
  \caption{Dependence on the $2500\;$\AA\ monochromatic luminosity of \pedap{\alpha}{ox}{$\prime$} ({\it upper panel}) and \pedix{L}{2\kev} ({\it lower panel}).
           Black filled circles mark data from \citet{strateva05} for their ``main'' \sdss\ sample ($155$ objects with $0.1\lesssim z\lesssim 4.5$); arrows indicate upper limits in the X-ray detection.
	   Dashed lines are the best--fit linear relations for their combined sample (``main'' \sdss\ sample $+$ high-redshift sample $+$ Seyfert~1 sample): $\pedap{\alpha}{ox}{$\prime$}=-0.136\cdot \pedix{\log L}{2500\ang}+2.616$ ({\it upper panel}) and $\pedix{\log L}{2\kev}=0.648\cdot \pedix{\log L}{2500\ang}+6.734$ ({\it lower panel}).
	   Red open star marks the position of \sorg\ as observed in \dataN\ (to avoid confusion we report only the values from the (A) data, being the two states indistinguishable).}
  \label{fig:luvVSstra}%
\end{figure}

Many authors have investigated the relation between rest-frame UV and soft X-ray AGN emission, and its dependence with redshift and/or optical luminosity.
Most studies have concluded that there is no evidence for a redshift dependence, while the X-ray emission results to be correlated with the UV emission, and the ratio of the monochromatic X-ray and UV luminosities \pedix{\alpha}{ox} decreases with increasing the latter \citep[\eg,][]{vignali03,strateva05,steffen06,just07}.
In particular, \citet{strateva05} combined radio quiet sources from the SDSS, a heterogeneous low-redshift Seyfert~1 sample, and a heterogeneous high-redshift sample.
Performing their analysis both with and without the Seyfert~1 sample, they found that the monochromatic luminosities at $2500\;$\AA\ and $2\;$keV are correlated (with a slope lower than $1$).
The broad-band spectral index, calculated here at $\lambda=2500\;$\AA, $\pedap{\alpha}{ox}{$\prime$} \equiv \log (\pedix{F}{2\kev}/\pedix{F}{2500\ang})/\log (\pedix{\nu}{2\kev}/\pedix{\nu}{2500\ang})$, results to be anticorrelated to the rest-frame monochromatic UV luminosity.
The \dataN\ \xmm\ observations put \sorg\ outside both relations (see Fig~\ref{fig:luvVSstra}).
From the observed luminosity at $2\;$keV we would expect a spectral index\footnote{The expected luminosities at $2500\;$\AA\ for the (A) and (B) spectra are $\pedix{L}{2500\ang}=1.65\times 10^{27}$ and $2\times 10^{27}\;$\lum~Hz$^{-1}$, respectively.} $\pedap{\alpha}{ox}{$\prime$}\sim -1.09$, while the UV luminosity obtained from the OM data would imply an index $\pedap{\alpha}{ox}{$\prime$}\sim -1.29$.
Both values differ considerably from the observed indices, $\pedap{\alpha}{ox}{$\prime$} = -1.66$ and $-1.67$ for the (A) and (B) spectra, respectively.
These numbers should be taken with caution: the host-galaxy contribution was not excluded from the \dataN\ OM data, thus the nuclear UV flux could be lower than the value estimated here. 
However, taking this consideration in mind, \sorg\ appears to be still underluminous in the X-ray band if compared with optically selected sources in the same bin of UV luminosity\footnote{From Fig.~10 in \citet{strateva05}, only $1$ of $37$ Seyfert~1 falls in this bin of $\pedap{\alpha}{ox}{$\prime$}$ ({\it upper panel}), while in their main SDSS sample they have only $1$ AGN with $\pedix{L}{2500\ang}<3.16\times10^{30}\;$\lum~Hz$^{-1}$ with $-1.7<\pedap{\alpha}{ox}{$\prime$}<-1.6$ (compared with $16$ sources with $\pedix{L}{2500\ang}>3.16\times10^{30}\;$\lum~Hz$^{-1}$, {\it lower panel}).}, a property shown by other NLS1s with high-energy spectral complexity \citep[\eg,][]{gallo06}.


\section{Conclusions}

\sorg\ is a bright Narrow Line Seyfert~1 galaxy, whose high energy emission shows strong variability both in shape and flux.
In this paper, we have reported a consistent analysis of the three available \xmm\ datasets of this object; the observations taken in \dataN\ are presented here for the first time.
The availability of high quality data obtained at different epochs with the same satellite opened us the very intriguing possibility to study in details the complex emission of \sorg, the origin of the spectral features as well as of the strong variability observed. 
The main aspects of this work can be summarized as follows.

   \begin{enumerate}
      \item Strong flux and spectral variability is observed on timescales of years and shorter; the strongest variation is observed in the soft X-ray band, at energies lower than $3\;$keV.
      The importance of absorption structures is evident in all the states, a result consistent with the strong polarization continuum and the absorption features observed in the optical-UV.
      \item The broad band ($0.4-10\;$keV) EPIC spectra can be well described either assuming an intrinsic continuum and reflection from an accretion disk covered by a system of ionized absorbing materials (the ``reflection-based'' model), or by an intrinsic continuum and a scattered component, with a complex combination of neutral absorbing material and a two-phase warm absorber covering the central source in different ways (the ``absorption-based'' model).
      In the former case, the \xmm\ observations imply evidence of important reprocessing from the disk of the intrinsic continuum (in terms of relativistically blurred reflection component and broad iron line) for all the three states.
      \item In both scenarios, the analysis suggests that the observed variability can be ascribed mainly to changes in the absorbers, although in the former  the ionization of the emitting region of the disk plays a role, too.
      Further deeper observations are needed to model correctly the continuum in the $4-7\;$keV range and better constrain the parameters of the broad line and/or the absorbing structures, a task made particularly difficult by the well known interplay between absorbers and broad features in the resulting spectrum.
      Moreover, a spectral coverage extended at higher energy would be able to fix with higher accuracy the intrinsic spectral index and the contribution from the reflection.
      \item Finally, unrelated X-ray and optical variabilities imply a {\it change in the classification of \sorg\ as X-ray weak source}. 
      The classification of a source as X-ray weak AGN seems to depend strongly on the date of the high energy observations; the suggestion is that a similar behaviour can occur also for other sources in intermediate range $-2 < \pedix{\alpha}{ox} < -1.7$.
      The nature of these sources as homogeneous class is not so clear, numbering among them objects like \object{PHL~1811}, that seems to be intrinsically X-ray weak \citep{leighly07ii,leighly07i}, or \object{PG~2112$+$059}, for which the weakness and the X-ray variability are ascribed to change in the intrinsic continuum instead of to absorption effects \citep{sch07}.
   \end{enumerate}
   
To conclude, whatever the model used, the changes observed in the emission of \sorg\ imply physical variations occurring in regions close to the central black hole.
Moreover, \pedix{\alpha}{ox} is definitely a variable parameter \citep[see also][]{gallo06,grupe06}: it seems that some sources go through phases where they appear to be X-ray weak and for most of them the X-ray weakness is a temporary event.
Further investigations of their high energy emission with a simultaneous multiwavelength coverage are needed.


\begin{acknowledgements}

  We warmly thank the referee, Dr. Grupe, for his suggestions that significantly improved the paper.
  We are grateful to T.~Kallman, C.~Gordon and K.~A.~Arnaud for enlightening discussions regarding XSTAR and XSPEC interplay.
  LB thanks L.~Maraschi and V.~Braito for useful discussions.
  MG would like to thank G.~Ponti, M.~Dadina and G.~Palumbo for key and pleasant discussions.
  MC, MG, and CV acknowledge finantial support from ASI contract number ASI/INAF I/023/05/0.
  Based on observations obtained with \xmm\ (an ESA science mission with instruments and contributions directly funded by ESA Member States and the USA, NASA).

\end{acknowledgements}

\bibliographystyle{aa} 
\bibliography{msPG1535.bib} 

\begin{thebibliography}{108}
\expandafter\ifx\csname natexlab\endcsname\relax\def\natexlab#1{#1}\fi

\bibitem[{{Arnaud}(1996)}]{xspec}
{Arnaud}, K.~A. 1996, in Astronomical Society of the Pacific Conference Series,
  Vol. 101, Astronomical Data Analysis Software and Systems V, ed. G.~H.
  {Jacoby} \& J.~{Barnes}, 17--+

\bibitem[{{Ashton} {et~al.}(2004){Ashton}, {Page}, {Blustin}, {Puchnarewicz},
  {Branduardi-Raymont}, {Mason}, {C{\'o}rdova}, \& {Priedhorsky}}]{ashton04}
{Ashton}, C.~E., {Page}, M.~J., {Blustin}, A.~J., {et~al.} 2004, \mnras, 355,
  73

\bibitem[{{Berriman} {et~al.}(1990){Berriman}, {Schmidt}, {West}, \&
  {Stockman}}]{berriman90}
{Berriman}, G., {Schmidt}, G.~D., {West}, S.~C., \& {Stockman}, H.~S. 1990,
  \apjs, 74, 869

\bibitem[{{Blustin} {et~al.}(2005){Blustin}, {Page}, {Fuerst},
  {Branduardi-Raymont}, \& {Ashton}}]{blustin05}
{Blustin}, A.~J., {Page}, M.~J., {Fuerst}, S.~V., {Branduardi-Raymont}, G., \&
  {Ashton}, C.~E. 2005, \aap, 431, 111

\bibitem[{{Boller} {et~al.}(2007){Boller}, {Balestra}, \&
  {Kollatschny}}]{boller07}
{Boller}, T., {Balestra}, I., \& {Kollatschny}, W. 2007, \aap, 465, 87

\bibitem[{{Boller} {et~al.}(2002){Boller}, {Fabian}, {Sunyaev}, {Tr{\"u}mper},
  {Vaughan}, {Ballantyne}, {Brandt}, {Keil}, \& {Iwasawa}}]{boller02}
{Boller}, T., {Fabian}, A.~C., {Sunyaev}, R., {et~al.} 2002, \mnras, 329, L1

\bibitem[{{Boroson}(2002)}]{boroson02}
{Boroson}, T.~A. 2002, \apj, 565, 78

\bibitem[{{Boroson} \& {Green}(1992)}]{boroson92}
{Boroson}, T.~A. \& {Green}, R.~F. 1992, \apjs, 80, 109

\bibitem[{{Brandt} {et~al.}(2000){Brandt}, {Laor}, \& {Wills}}]{brandt00}
{Brandt}, W.~N., {Laor}, A., \& {Wills}, B.~J. 2000, \apj, 528, 637

\bibitem[{{Brinkmann} {et~al.}(2003){Brinkmann}, {Grupe}, {Branduardi-Raymont},
  \& {Ferrero}}]{brinkmann03}
{Brinkmann}, W., {Grupe}, D., {Branduardi-Raymont}, G., \& {Ferrero}, E. 2003,
  \aap, 398, 81

\bibitem[{{Brinkmann} {et~al.}(2004){Brinkmann}, {Papadakis}, \&
  {Ferrero}}]{brinkmann04}
{Brinkmann}, W., {Papadakis}, I.~E., \& {Ferrero}, E. 2004, \aap, 414, 107

\bibitem[{{Cappi}(2006)}]{cappi06}
{Cappi}, M. 2006, Astronomische Nachrichten, 327, 1012

\bibitem[{{Carini} {et~al.}(2007){Carini}, {Noble}, {Taylor}, \&
  {Culler}}]{carini07}
{Carini}, M.~T., {Noble}, J.~C., {Taylor}, R., \& {Culler}, R. 2007, \aj, 133,
  303

\bibitem[{{Chevallier} {et~al.}(2006){Chevallier}, {Collin}, {Dumont},
  {Czerny}, {Mouchet}, {Gon{\c c}alves}, \& {Goosmann}}]{chevallier06}
{Chevallier}, L., {Collin}, S., {Dumont}, A.-M., {et~al.} 2006, \aap, 449, 493

\bibitem[{{Costantini} {et~al.}(2007){Costantini}, {Kaastra}, {Arav}, {Kriss},
  {Steenbrugge}, {Gabel}, {Verbunt}, {Behar}, {Gaskell}, {Korista}, {Proga},
  {Quijano}, {Scott}, {Klimek}, \& {Hedrick}}]{costantini07}
{Costantini}, E., {Kaastra}, J.~S., {Arav}, N., {et~al.} 2007, \aap, 461, 121

\bibitem[{{Crummy} {et~al.}(2006){Crummy}, {Fabian}, {Gallo}, \&
  {Ross}}]{crummy06}
{Crummy}, J., {Fabian}, A.~C., {Gallo}, L., \& {Ross}, R.~R. 2006, \mnras, 365,
  1067

\bibitem[{{Dadina} \& {Cappi}(2004)}]{dadina04}
{Dadina}, M. \& {Cappi}, M. 2004, \aap, 413, 921

\bibitem[{{De Vaucouleurs} {et~al.}(1991){De Vaucouleurs}, {De Vaucouleurs},
  {Corwin Jr.}, {Buta}, {Paturel}, \& {Fouque}}]{red}
{De Vaucouleurs}, G., {De Vaucouleurs}, A., {Corwin Jr.}, H.~G., {et~al.} 1991,
  Springer-Verlag: New York

\bibitem[{{de Veny} \& {Lynds}(1969)}]{deveny69}
{de Veny}, J.~B. \& {Lynds}, C.~R. 1969, \pasp, 81, 535

\bibitem[{{den Herder} {et~al.}(2001){den Herder}, {Brinkman}, {Kahn},
  {Branduardi-Raymont}, {Thomsen}, {Aarts}, {Audard}, {Bixler}, {den Boggende},
  {Cottam}, {Decker}, {Dubbeldam}, {Erd}, {Goulooze}, {G{\"u}del}, {Guttridge},
  {Hailey}, {Janabi}, {Kaastra}, {de Korte}, {van Leeuwen}, {Mauche},
  {McCalden}, {Mewe}, {Naber}, {Paerels}, {Peterson}, {Rasmussen}, {Rees},
  {Sakelliou}, {Sako}, {Spodek}, {Stern}, {Tamura}, {Tandy}, {de Vries},
  {Welch}, \& {Zehnder}}]{rgs}
{den Herder}, J.~W., {Brinkman}, A.~C., {Kahn}, S.~M., {et~al.} 2001, \aap,
  365, L7

\bibitem[{{Ehle} {et~al.}(2001){Ehle}, {Breitfellner}, {Dahlem}, {Guainazzi},
  {Rodriguez}, {Santos-Lleo}, {Schartel}, \& {Tomas}}]{xmmhb}
{Ehle}, M., {Breitfellner}, M., {Dahlem}, M., {et~al.} 2001, \xmm\ Users'
  Handbook Issue 2.0
  (http://xmm.vilspa.esa.es/external/xmm\_user\_support/\\documentation/uhb\_2%
.0/index.html)

\bibitem[{{Fabian} {et~al.}(2004){Fabian}, {Miniutti}, {Gallo}, {Boller},
  {Tanaka}, {Vaughan}, \& {Ross}}]{fabian04}
{Fabian}, A.~C., {Miniutti}, G., {Gallo}, L., {et~al.} 2004, \mnras, 353, 1071

\bibitem[{{Fabian} {et~al.}(2005){Fabian}, {Miniutti}, {Iwasawa}, \&
  {Ross}}]{fabian05}
{Fabian}, A.~C., {Miniutti}, G., {Iwasawa}, K., \& {Ross}, R.~R. 2005, \mnras,
  361, 795

\bibitem[{{Fabian} {et~al.}(2002){Fabian}, {Vaughan}, {Nandra}, {Iwasawa},
  {Ballantyne}, {Lee}, {De Rosa}, {Turner}, \& {Young}}]{fabian02}
{Fabian}, A.~C., {Vaughan}, S., {Nandra}, K., {et~al.} 2002, \mnras, 335, L1

\bibitem[{{Gallagher} {et~al.}(2001){Gallagher}, {Brandt}, {Laor}, {Elvis},
  {Mathur}, {Wills}, \& {Iyomoto}}]{gallagher01}
{Gallagher}, S.~C., {Brandt}, W.~N., {Laor}, A., {et~al.} 2001, \apj, 546, 795

\bibitem[{{Gallo}(2006)}]{gallo06}
{Gallo}, L.~C. 2006, \mnras, 368, 479

\bibitem[{{Gallo} {et~al.}(2004){Gallo}, {Tanaka}, {Boller}, {Fabian},
  {Vaughan}, \& {Brandt}}]{gallo04}
{Gallo}, L.~C., {Tanaka}, Y., {Boller}, T., {et~al.} 2004, \mnras, 353, 1064

\bibitem[{{Gierli{\'n}ski} \& {Done}(2004)}]{swind04}
{Gierli{\'n}ski}, M. \& {Done}, C. 2004, \mnras, 349, L7

\bibitem[{{Gierli{\'n}ski} \& {Done}(2006)}]{swind06}
{Gierli{\'n}ski}, M. \& {Done}, C. 2006, \mnras, 371, L16

\bibitem[{{Grupe} {et~al.}(1995){Grupe}, {Beuerman}, {Mannheim}, {Thomas},
  {Fink}, \& {de Martino}}]{grupe95}
{Grupe}, D., {Beuerman}, K., {Mannheim}, K., {et~al.} 1995, \aap, 300, L21+

\bibitem[{{Grupe} {et~al.}(2007){Grupe}, {Komossa}, \& {Gallo}}]{grupe07}
{Grupe}, D., {Komossa}, S., \& {Gallo}, L.~C. 2007, \apjl, 668, L111

\bibitem[{{Grupe} {et~al.}(2006){Grupe}, {Leighly}, {Komossa}, {Schady},
  {O'Brien}, {Burrows}, \& {Nousek}}]{grupe06}
{Grupe}, D., {Leighly}, K.~M., {Komossa}, S., {et~al.} 2006, \aj, 132, 1189

\bibitem[{{Grupe} \& {Mathur}(2004)}]{grupe04nlsy1}
{Grupe}, D. \& {Mathur}, S. 2004, \apjl, 606, L41

\bibitem[{{Grupe} {et~al.}(2004{\natexlab{a}}){Grupe}, {Mathur}, \&
  {Komossa}}]{grupe04abs}
{Grupe}, D., {Mathur}, S., \& {Komossa}, S. 2004{\natexlab{a}}, \aj, 127, 3161

\bibitem[{{Grupe} {et~al.}(2004{\natexlab{b}}){Grupe}, {Wills}, {Leighly}, \&
  {Meusinger}}]{grupe04}
{Grupe}, D., {Wills}, B.~J., {Leighly}, K.~M., \& {Meusinger}, H.
  2004{\natexlab{b}}, \aj, 127, 156

\bibitem[{{Guainazzi} \& {Bianchi}(2007)}]{guainazzi07}
{Guainazzi}, M. \& {Bianchi}, S. 2007, \mnras, 374, 1290

\bibitem[{{Guainazzi} {et~al.}(2006){Guainazzi}, {Bianchi}, \& {Dov{\v
  c}iak}}]{guainazzi06}
{Guainazzi}, M., {Bianchi}, S., \& {Dov{\v c}iak}, M. 2006, Astronomische
  Nachrichten, 327, 1032

\bibitem[{{Haas} {et~al.}(2003){Haas}, {Klaas}, {M{\"u}ller}, {Bertoldi},
  {Camenzind}, {Chini}, {Krause}, {Lemke}, {Meisenheimer}, {Richards}, \&
  {Wilkes}}]{haas03}
{Haas}, M., {Klaas}, U., {M{\"u}ller}, S.~A.~H., {et~al.} 2003, \aap, 402, 87

\bibitem[{{Halpern}(1984)}]{halpern84}
{Halpern}, J.~P. 1984, \apj, 281, 90

\bibitem[{{Hill} {et~al.}(1996){Hill}, {Goodrich}, \& {Depoy}}]{hill96}
{Hill}, G.~J., {Goodrich}, R.~W., \& {Depoy}, D.~L. 1996, \apj, 462, 163

\bibitem[{{Immler} {et~al.}(2003){Immler}, {Brandt}, {Vignali}, {Bauer},
  {Crenshaw}, {Feldmeier}, \& {Kraemer}}]{immler03}
{Immler}, S., {Brandt}, W.~N., {Vignali}, C., {et~al.} 2003, \aj, 126, 153

\bibitem[{{Jim{\'e}nez-Bail{\'o}n} {et~al.}(2005){Jim{\'e}nez-Bail{\'o}n},
  {Piconcelli}, {Guainazzi}, {Schartel}, {Rodr{\'{\i}}guez-Pascual}, \&
  {Santos-Lle{\'o}}}]{elena05}
{Jim{\'e}nez-Bail{\'o}n}, E., {Piconcelli}, E., {Guainazzi}, M., {et~al.} 2005,
  \aap, 435, 449

\bibitem[{{Just} {et~al.}(2007){Just}, {Brandt}, {Shemmer}, {Steffen},
  {Schneider}, {Chartas}, \& {Garmire}}]{just07}
{Just}, D.~W., {Brandt}, W.~N., {Shemmer}, O., {et~al.} 2007, \apj, 665, 1004

\bibitem[{{Kallman} \& {Bautista}(2001)}]{xstar}
{Kallman}, T. \& {Bautista}, M. 2001, \apjs, 133, 221

\bibitem[{{Kellermann} {et~al.}(1989){Kellermann}, {Sramek}, {Schmidt},
  {Shaffer}, \& {Green}}]{kellermann89}
{Kellermann}, K.~I., {Sramek}, R., {Schmidt}, M., {Shaffer}, D.~B., \& {Green},
  R. 1989, \aj, 98, 1195

\bibitem[{{Kobayashi} {et~al.}(1993){Kobayashi}, {Sato}, {Yamashita}, {Shiba},
  \& {Takami}}]{kobayashi93}
{Kobayashi}, Y., {Sato}, S., {Yamashita}, T., {Shiba}, H., \& {Takami}, H.
  1993, \apj, 404, 94

\bibitem[{{Laor}(1991)}]{laor91}
{Laor}, A. 1991, \apj, 376, 90

\bibitem[{{Laor} {et~al.}(1997){Laor}, {Fiore}, {Elvis}, {Wilkes}, \&
  {McDowell}}]{laor97}
{Laor}, A., {Fiore}, F., {Elvis}, M., {Wilkes}, B.~J., \& {McDowell}, J.~C.
  1997, \apj, 477, 93

\bibitem[{{Leighly}(1999)}]{leighly99}
{Leighly}, K.~M. 1999, \apjs, 125, 317

\bibitem[{{Leighly} {et~al.}(2007{\natexlab{a}}){Leighly}, {Halpern},
  {Jenkins}, \& {Casebeer}}]{leighly07ii}
{Leighly}, K.~M., {Halpern}, J.~P., {Jenkins}, E.~B., \& {Casebeer}, D.
  2007{\natexlab{a}}, \apjs, 173, 1

\bibitem[{{Leighly} {et~al.}(2007{\natexlab{b}}){Leighly}, {Halpern},
  {Jenkins}, {Grupe}, {Choi}, \& {Prescott}}]{leighly07i}
{Leighly}, K.~M., {Halpern}, J.~P., {Jenkins}, E.~B., {et~al.}
  2007{\natexlab{b}}, \apj, 663, 103

\bibitem[{{Longinotti} {et~al.}(2007){Longinotti}, {Sim}, {Nandra}, \&
  {Cappi}}]{longinotti07}
{Longinotti}, A.~L., {Sim}, S.~A., {Nandra}, K., \& {Cappi}, M. 2007, \mnras,
  374, 237

\bibitem[{{Magdziarz} \& {Zdziarski}(1995)}]{pexrav}
{Magdziarz}, P. \& {Zdziarski}, A.~A. 1995, \mnras, 273, 837

\bibitem[{{Makishima}(1986)}]{makishima86}
{Makishima}, K. 1986, in Lecture Notes in Physics, Berlin Springer Verlag, Vol.
  266, The Physics of Accretion onto Compact Objects, ed. K.~O. {Mason}, M.~G.
  {Watson}, \& N.~E. {White}, 249--+

\bibitem[{{Martocchia} {et~al.}(2002){Martocchia}, {Matt}, \&
  {Karas}}]{martocchia02}
{Martocchia}, A., {Matt}, G., \& {Karas}, V. 2002, \aap, 383, L23

\bibitem[{{Mason} {et~al.}(2001){Mason}, {Breeveld}, {Much}, {Carter},
  {Cordova}, {Cropper}, {Fordham}, {Huckle}, {Ho}, {Kawakami}, {Kennea},
  {Kennedy}, {Mittaz}, {Pandel}, {Priedhorsky}, {Sasseen}, {Shirey}, {Smith},
  \& {Vreux}}]{om}
{Mason}, K.~O., {Breeveld}, A., {Much}, R., {et~al.} 2001, \aap, 365, L36

\bibitem[{{Mathur} \& {Grupe}(2005)}]{mathur05}
{Mathur}, S. \& {Grupe}, D. 2005, \aap, 432, 463

\bibitem[{{McKernan} {et~al.}(2007){McKernan}, {Yaqoob}, \&
  {Reynolds}}]{mckernan07}
{McKernan}, B., {Yaqoob}, T., \& {Reynolds}, C.~S. 2007, \mnras, 379, 1359

\bibitem[{{Merloni} {et~al.}(2006){Merloni}, {Malzac}, {Fabian}, \&
  {Ross}}]{merloni06}
{Merloni}, A., {Malzac}, J., {Fabian}, A.~C., \& {Ross}, R.~R. 2006, \mnras,
  370, 1699

\bibitem[{{Miller} {et~al.}(2006){Miller}, {Turner}, {Reeves}, {George},
  {Porquet}, {Nandra}, \& {Dovciak}}]{miller06}
{Miller}, L., {Turner}, T.~J., {Reeves}, J.~N., {et~al.} 2006, \aap, 453, L13

\bibitem[{{Miniutti} \& {Fabian}(2004)}]{miniutti04}
{Miniutti}, G. \& {Fabian}, A.~C. 2004, \mnras, 349, 1435

\bibitem[{{Miniutti} {et~al.}(2003){Miniutti}, {Fabian}, {Goyder}, \&
  {Lasenby}}]{miniutti03}
{Miniutti}, G., {Fabian}, A.~C., {Goyder}, R., \& {Lasenby}, A.~N. 2003,
  \mnras, 344, L22

\bibitem[{{Miniutti} {et~al.}(2007){Miniutti}, {Ponti}, {Dadina}, {Cappi}, \&
  {Malaguti}}]{miniutti07}
{Miniutti}, G., {Ponti}, G., {Dadina}, M., {Cappi}, M., \& {Malaguti}, G. 2007,
  \mnras, 375, 227

\bibitem[{{Murphy} {et~al.}(1996){Murphy}, {Lockman}, {Laor}, \& {Elvis}}]{nh}
{Murphy}, E.~M., {Lockman}, F.~J., {Laor}, A., \& {Elvis}, M. 1996, \apjs, 105,
  369

\bibitem[{{Nandra} {et~al.}(2007){Nandra}, {O'Neill}, {George}, \&
  {Reeves}}]{nandra07}
{Nandra}, K., {O'Neill}, P.~M., {George}, I.~M., \& {Reeves}, J.~N. 2007,
  \mnras, 942

\bibitem[{{Narayanan} {et~al.}(2004){Narayanan}, {Hamann}, {Barlow},
  {Burbidge}, {Cohen}, {Junkkarinen}, \& {Lyons}}]{narayanan04}
{Narayanan}, D., {Hamann}, F., {Barlow}, T., {et~al.} 2004, \apj, 601, 715

\bibitem[{{Netzer} \& {Laor}(1993)}]{netzer93}
{Netzer}, H. \& {Laor}, A. 1993, \apjl, 404, L51

\bibitem[{{Neugebauer} {et~al.}(1987){Neugebauer}, {Green}, {Matthews},
  {Schmidt}, {Soifer}, \& {Bennett}}]{neugebauer87}
{Neugebauer}, G., {Green}, R.~F., {Matthews}, K., {et~al.} 1987, \apjs, 63, 615

\bibitem[{{Neugebauer} \& {Matthews}(1999)}]{neugebauer99}
{Neugebauer}, G. \& {Matthews}, K. 1999, \aj, 118, 35

\bibitem[{{Nicastro} {et~al.}(1999){Nicastro}, {Fiore}, \& {Matt}}]{nicastro99}
{Nicastro}, F., {Fiore}, F., \& {Matt}, G. 1999, \apj, 517, 108

\bibitem[{{O'Neill} {et~al.}(2007){O'Neill}, {Nandra}, {Cappi}, {Longinotti},
  \& {Sim}}]{oneill07}
{O'Neill}, P.~M., {Nandra}, K., {Cappi}, M., {Longinotti}, A.~L., \& {Sim},
  S.~A. 2007, \mnras, 381, L94

\bibitem[{{Osterbrock} \& {Pogge}(1987)}]{osterbrock87}
{Osterbrock}, D.~E. \& {Pogge}, R.~W. 1987, \apj, 323, 108

\bibitem[{{Petrucci} {et~al.}(2007){Petrucci}, {Ponti}, {Matt}, {Longinotti},
  {Malzac}, {Mouchet}, {Boisson}, {Maraschi}, {Nandra}, \&
  {Ferrando}}]{petrucci07}
{Petrucci}, P.~O., {Ponti}, G., {Matt}, G., {et~al.} 2007, \aap, 470, 889

\bibitem[{{Phillips}(1978)}]{phillips78}
{Phillips}, M.~M. 1978, \apjs, 38, 187

\bibitem[{{Piconcelli} {et~al.}(2004){Piconcelli}, {Jimenez-Bail{\'o}n},
  {Guainazzi}, {Schartel}, {Rodr{\'{\i}}guez-Pascual}, \&
  {Santos-Lle{\'o}}}]{piconcelli04}
{Piconcelli}, E., {Jimenez-Bail{\'o}n}, E., {Guainazzi}, M., {et~al.} 2004,
  \mnras, 351, 161

\bibitem[{{Piconcelli} {et~al.}(2005){Piconcelli}, {Jimenez-Bail{\'o}n},
  {Guainazzi}, {Schartel}, {Rodr{\'{\i}}guez-Pascual}, \&
  {Santos-Lle{\'o}}}]{piconcelli05}
{Piconcelli}, E., {Jimenez-Bail{\'o}n}, E., {Guainazzi}, M., {et~al.} 2005,
  \aap, 432, 15

\bibitem[{{Pounds} {et~al.}(2004){Pounds}, {Reeves}, {Page}, \&
  {O'Brien}}]{pounds04}
{Pounds}, K.~A., {Reeves}, J.~N., {Page}, K.~L., \& {O'Brien}, P.~T. 2004,
  \apj, 616, 696

\bibitem[{{Proga} \& {Kallman}(2004)}]{proga04}
{Proga}, D. \& {Kallman}, T.~R. 2004, \apj, 616, 688

\bibitem[{{Protassov} {et~al.}(2002){Protassov}, {van Dyk}, {Connors},
  {Kashyap}, \& {Siemiginowska}}]{protassov02}
{Protassov}, R., {van Dyk}, D.~A., {Connors}, A., {Kashyap}, V.~L., \&
  {Siemiginowska}, A. 2002, \apj, 571, 545

\bibitem[{{Reeves} {et~al.}(2004){Reeves}, {Nandra}, {George}, {Pounds},
  {Turner}, \& {Yaqoob}}]{reeves04}
{Reeves}, J.~N., {Nandra}, K., {George}, I.~M., {et~al.} 2004, \apj, 602, 648

\bibitem[{{Reynolds} \& {Nowak}(2003)}]{reynolds03}
{Reynolds}, C.~S. \& {Nowak}, M.~A. 2003, \physrep, 377, 389

\bibitem[{{Risaliti} {et~al.}(2003){Risaliti}, {Elvis}, {Gilli}, \&
  {Salvati}}]{risaliti03}
{Risaliti}, G., {Elvis}, M., {Gilli}, R., \& {Salvati}, M. 2003, \apjl, 587, L9

\bibitem[{{Ross} \& {Fabian}(2005)}]{ross05}
{Ross}, R.~R. \& {Fabian}, A.~C. 2005, \mnras, 358, 211

\bibitem[{{Ross} {et~al.}(1999){Ross}, {Fabian}, \& {Young}}]{ross99}
{Ross}, R.~R., {Fabian}, A.~C., \& {Young}, A.~J. 1999, \mnras, 306, 461

\bibitem[{{Schartel} {et~al.}(2007){Schartel}, {Rodr{\'{\i}}guez-Pascual},
  {Santos-Lle{\'o}}, {Ballo}, {Clavel}, {Guainazzi}, {Jim{\'e}nez-Bail{\'o}n},
  \& {Piconcelli}}]{sch07}
{Schartel}, N., {Rodr{\'{\i}}guez-Pascual}, P.~M., {Santos-Lle{\'o}}, M.,
  {et~al.} 2007, \aap, 474, 431

\bibitem[{{Schartel} {et~al.}(2005){Schartel}, {Rodr{\'{\i}}guez-Pascual},
  {Santos-Lle{\'o}}, {Clavel}, {Guainazzi}, {Jim{\'e}nez-Bail{\'o}n}, \&
  {Piconcelli}}]{sch05}
{Schartel}, N., {Rodr{\'{\i}}guez-Pascual}, P.~M., {Santos-Lle{\'o}}, M.,
  {et~al.} 2005, \aap, 433, 455

\bibitem[{{Schmidt} \& {Green}(1983)}]{schmidt83}
{Schmidt}, M. \& {Green}, R.~F. 1983, \apj, 269, 352

\bibitem[{{Schurch} \& {Done}(2006)}]{schurch06}
{Schurch}, N.~J. \& {Done}, C. 2006, \mnras, 371, 81

\bibitem[{{Shuder} \& {Osterbrock}(1981)}]{shuder81}
{Shuder}, J.~M. \& {Osterbrock}, D.~E. 1981, \apj, 250, 55

\bibitem[{{Smith} {et~al.}(2004){Smith}, {Robinson}, {Alexander}, {Young},
  {Axon}, \& {Corbett}}]{smith04}
{Smith}, J.~E., {Robinson}, A., {Alexander}, D.~M., {et~al.} 2004, \mnras, 350,
  140

\bibitem[{{Smith} {et~al.}(2002){Smith}, {Young}, {Robinson}, {Corbett},
  {Giannuzzo}, {Axon}, \& {Hough}}]{smith02}
{Smith}, J.~E., {Young}, S., {Robinson}, A., {et~al.} 2002, \mnras, 335, 773

\bibitem[{{Smith} {et~al.}(1997){Smith}, {Schmidt}, {Allen}, \&
  {Hines}}]{smith97}
{Smith}, P.~S., {Schmidt}, G.~D., {Allen}, R.~G., \& {Hines}, D.~C. 1997, \apj,
  488, 202

\bibitem[{{Sobolewska} \& {Done}(2007)}]{sobolewska07}
{Sobolewska}, M.~A. \& {Done}, C. 2007, \mnras, 374, 150

\bibitem[{{Steffen} {et~al.}(2006){Steffen}, {Strateva}, {Brandt}, {Alexander},
  {Koekemoer}, {Lehmer}, {Schneider}, \& {Vignali}}]{steffen06}
{Steffen}, A.~T., {Strateva}, I., {Brandt}, W.~N., {et~al.} 2006, \aj, 131,
  2826

\bibitem[{{Strateva} {et~al.}(2005){Strateva}, {Brandt}, {Schneider}, {Vanden
  Berk}, \& {Vignali}}]{strateva05}
{Strateva}, I.~V., {Brandt}, W.~N., {Schneider}, D.~P., {Vanden Berk}, D.~G.,
  \& {Vignali}, C. 2005, \aj, 130, 387

\bibitem[{{Str{\"u}der} {et~al.}(2001){Str{\"u}der}, {Briel}, {Dennerl},
  {Hartmann}, {Kendziorra}, {Meidinger}, {Pfeffermann}, {Reppin}, {Aschenbach},
  {Bornemann}, {Br{\"a}uninger}, {Burkert}, {Elender}, {Freyberg}, {Haberl},
  {Hartner}, {Heuschmann}, {Hippmann}, {Kastelic}, {Kemmer}, {Kettenring},
  {Kink}, {Krause}, {M{\"u}ller}, {Oppitz}, {Pietsch}, {Popp}, {Predehl},
  {Read}, {Stephan}, {St{\"o}tter}, {Tr{\"u}mper}, {Holl}, {Kemmer}, {Soltau},
  {St{\"o}tter}, {Weber}, {Weichert}, {von Zanthier}, {Carathanassis}, {Lutz},
  {Richter}, {Solc}, {B{\"o}ttcher}, {Kuster}, {Staubert}, {Abbey}, {Holland},
  {Turner}, {Balasini}, {Bignami}, {La Palombara}, {Villa}, {Buttler},
  {Gianini}, {Lain{\'e}}, {Lumb}, \& {Dhez}}]{pn}
{Str{\"u}der}, L., {Briel}, U., {Dennerl}, K., {et~al.} 2001, \aap, 365, L18

\bibitem[{{Sulentic} {et~al.}(2006){Sulentic}, {Dultzin-Hacyan}, {Marziani},
  {Bongardo}, {Braito}, {Calvani}, \& {Zamanov}}]{sulentic06}
{Sulentic}, J.~W., {Dultzin-Hacyan}, D., {Marziani}, P., {et~al.} 2006, Revista
  Mexicana de Astronomia y Astrofisica, 42, 23

\bibitem[{{Sulentic} {et~al.}(2000){Sulentic}, {Zwitter}, {Marziani}, \&
  {Dultzin-Hacyan}}]{sulentic00}
{Sulentic}, J.~W., {Zwitter}, T., {Marziani}, P., \& {Dultzin-Hacyan}, D. 2000,
  \apjl, 536, L5

\bibitem[{{Tanaka} {et~al.}(2004){Tanaka}, {Boller}, {Gallo}, {Keil}, \&
  {Ueda}}]{tanaka04}
{Tanaka}, Y., {Boller}, T., {Gallo}, L., {Keil}, R., \& {Ueda}, Y. 2004, \pasj,
  56, L9

\bibitem[{{Tremaine} {et~al.}(2002){Tremaine}, {Gebhardt}, {Bender}, {Bower},
  {Dressler}, {Faber}, {Filippenko}, {Green}, {Grillmair}, {Ho}, {Kormendy},
  {Lauer}, {Magorrian}, {Pinkney}, \& {Richstone}}]{tremaine02}
{Tremaine}, S., {Gebhardt}, K., {Bender}, R., {et~al.} 2002, \apj, 574, 740

\bibitem[{{Turner} {et~al.}(2001){Turner}, {Abbey}, {Arnaud}, {Balasini},
  {Barbera}, {Belsole}, {Bennie}, {Bernard}, {Bignami}, {Boer}, {Briel},
  {Butler}, {Cara}, {Chabaud}, {Cole}, {Collura}, {Conte}, {Cros}, {Denby},
  {Dhez}, {Di Coco}, {Dowson}, {Ferrando}, {Ghizzardi}, {Gianotti}, {Goodall},
  {Gretton}, {Griffiths}, {Hainaut}, {Hochedez}, {Holland}, {Jourdain},
  {Kendziorra}, {Lagostina}, {Laine}, {La Palombara}, {Lortholary}, {Lumb},
  {Marty}, {Molendi}, {Pigot}, {Poindron}, {Pounds}, {Reeves}, {Reppin},
  {Rothenflug}, {Salvetat}, {Sauvageot}, {Schmitt}, {Sembay}, {Short},
  {Spragg}, {Stephen}, {Str{\"u}der}, {Tiengo}, {Trifoglio}, {Tr{\"u}mper},
  {Vercellone}, {Vigroux}, {Villa}, {Ward}, {Whitehead}, \& {Zonca}}]{mos}
{Turner}, M.~J.~L., {Abbey}, A., {Arnaud}, M., {et~al.} 2001, \aap, 365, L27

\bibitem[{{V{\'e}ron-Cetty} {et~al.}(2001){V{\'e}ron-Cetty}, {V{\'e}ron}, \&
  {Gon{\c c}alves}}]{veron01}
{V{\'e}ron-Cetty}, M.-P., {V{\'e}ron}, P., \& {Gon{\c c}alves}, A.~C. 2001,
  \aap, 372, 730

\bibitem[{{Vestergaard} \& {Peterson}(2006)}]{vestergaard06}
{Vestergaard}, M. \& {Peterson}, B.~M. 2006, \apj, 641, 689

\bibitem[{{Vignali} {et~al.}(2003){Vignali}, {Brandt}, \&
  {Schneider}}]{vignali03}
{Vignali}, C., {Brandt}, W.~N., \& {Schneider}, D.~P. 2003, \aj, 125, 433

\bibitem[{{Weymann} {et~al.}(1991){Weymann}, {Morris}, {Foltz}, \&
  {Hewett}}]{weymann91}
{Weymann}, R.~J., {Morris}, S.~L., {Foltz}, C.~B., \& {Hewett}, P.~C. 1991,
  \apj, 373, 23

\bibitem[{{Zdziarski} {et~al.}(1995){Zdziarski}, {Johnson}, {Done}, {Smith}, \&
  {McNaron-Brown}}]{absori}
{Zdziarski}, A.~A., {Johnson}, W.~N., {Done}, C., {Smith}, D., \&
  {McNaron-Brown}, K. 1995, \apjl, 438, L63

\bibitem[{{Zhang} \& {Wang}(2006)}]{zhang06}
{Zhang}, E.-P. \& {Wang}, J.-M. 2006, \apj, 653, 137

\bibitem[{{Zhou} {et~al.}(2007){Zhou}, {Yang}, {L{\"u}}, \& {Wang}}]{zhou07}
{Zhou}, X.-L., {Yang}, F., {L{\"u}}, X.-R., \& {Wang}, J.-M. 2007, \aj, 133,
  432

\end{thebibliography}

\listofobjects

\end{document}